\documentclass[11pt,a4paper]{article}
\usepackage{jheppub}

\preprint{LU TP 14-43, MCNET-14-26}

\usepackage{def}

\usepackage{epstopdf} 
\usepackage{epsfig}
\usepackage{psfrag}
\usepackage{bbm}
\usepackage{color}

\usepackage{graphics}
\usepackage{axodraw}
\usepackage{inputenc}
\usepackage{xspace}
\usepackage{amsmath}
\usepackage{amssymb}
\usepackage{url}
\usepackage{rotating}

\skip\footins = 1\bigskipamount plus 2pt minus 4pt                              

\title{ColorFull -- a C++ library for calculations in SU(Nc) color space }
\author{Malin Sj\"odahl}
\emailAdd{malin.sjodahl@thep.lu.se}
\affiliation{Department of Astronomy and Theoretical Physics,\\
S\"olvegatan 14A, Lund University, SE-22362 Lund, Sweden}

\abstract{\ColorFull, a C++ package for treating QCD color structure,
is presented. \ColorFull, which utilizes the trace basis  
approach, is intended for interfacing with event 
generators, but can also be used as a stand-alone package
for squaring QCD amplitudes, calculating interferences, and
describing the effect of gluon emission and gluon exchange.

}

\begin{document}

\maketitle

\sloppy

\section{Introduction}
\label{sec:introduction}

The description of QCD color structure in the presence of many
external colored partons is a field of increased importance.
Some methods for performing automatic color summations of fully
contracted vacuum bubbles, for example as implemented in 
\texttt{FeynCalc} \cite{Mertig:1990an}, in the C program
\COLOR\, \cite{Hakkinen:1996bb}, or as presented in \cite{vanRitbergen:1998pn}, 
have been around for a while, and recently a more flexible 
Mathematica package, \ColorMath\,\cite{Sjodahl:2012nk}, 
allowing color structures with any number of open indices, has been published. 
Yet other general purpose event generator codes, such as 
\MadGraph\, \cite{Alwall:2011uj},
have separate built in routines for dealing with the color structure.

In the present paper a stand-alone C++ code, \ColorFull, designed 
for dealing with color contraction using color bases is presented\footnote{\ColorFull\, can be downloaded from 
\url{http://colorfull.hepforge.org/}.}.
\ColorFull\, is written with interfacing to event generators
in mind, and is currently interfaced to \Herwig\,(2.7)
\cite{Bahr:2008pv,Bellm:2013lba}, 
but can also be used as a stand-alone package for
investigations in color space.

\ColorFull\, is based on trace bases \cite{Paton:1969je,Dittner:1972hm,Cvi76,Cvitanovic:1980bu,Berends:1987cv,Mangano:1987xk,Mangano:1988kk,Nagy:2007ty,Sjodahl:2009wx}, 
where the basis vectors 
are given by (products of) closed and open quark-lines, but the 
code also offers functionality for reading in and treating any 
(orthogonal) basis for color space, such as multiplet bases 
\cite{Keppeler:2012ih, DecompositionPaper}.

The intent of this paper is to convey the underlying idea 
of \ColorFull. For full technical details we refer to the online Doxygen 
documentation\footnote{The automatically generated Doxygen documentation 
is available at \\ \hspace*{0.5 cm}\url{http://colorfull.hepforge.org/doxygen}.}. 
To set the stage, a brief introduction to the QCD color 
space is given in \secref{sec:color_space} and the trace
basis approach is presented in \secref{sec:trace_bases}.
Following this, some remarks about the computational strategy
are made in \secref{sec:strategy} and the key design 
features are presented in \secref{sec:colorfull}, whereas
examples of stand-alone usage are given in \secref{sec:standalone},
and  the interface to \Herwig\, is commented upon in \secref{sec:matchbox}.
The following section, \secref{sec:code}, describes the classes of 
\ColorFull\, and code validation is discussed in \secref{sec:validation}.
Finally some concluding remarks are made in \secref{sec:conclusions}.

\section{Color space}
\label{sec:color_space}

Apart from four-gluon vertices for which the color structure 
can be rewritten in terms of (one-gluon contracted) triple-gluon 
vertices, the QCD Lagrangian contains

\begin{equation}
\label{eq:t_and_f_def}
  \mbox{quark-gluon vertices,} \qquad 
  \epsfig{file=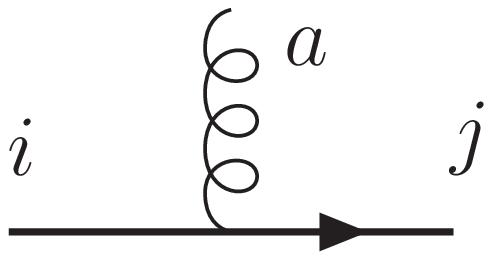,width=2.5cm} = (t^a)^i_{\phantom{i}j},
\end{equation} 
and
\begin{equation}
  \mbox{triple-gluon vertices,} \qquad 
  \parbox{2.5cm}{\epsfig{file=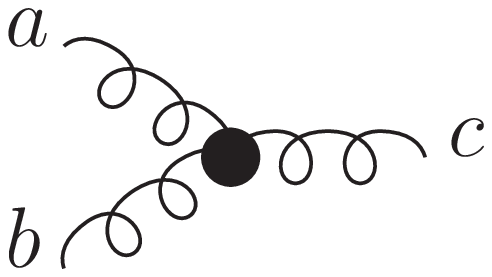,width=2.5cm}} 
  = i f^{abc},
\end{equation}
where we follow the convention of reading the fully anti-symmetric 
structure constant indices in counter clock-wise order.
The color structure of any amplitude, tree-level or beyond, pure QCD
or not, can thus be expressed in terms of these objects alone.
For observables we are -- as QCD is confining -- only interested
in color summed/averaged quantities.

Letting $\Col_1$ denote the color structure of the amplitude
under consideration, we are thus interested in $|\Col_1|^2$
where the scalar product\footnote{It is not hard to prove that this actually is a scalar product.}
is given by summing over all external color indices, i.e.,
\begin{equation}
  \left\langle \Col_1 | \Col_2 \right\rangle
  =\sum_{a_1,\,a_2,\,...}\Col_1^{* a_1\,a_2...} \, \Col_2^{a_1\,a_2...} 
\label{eq:scalar_product}
\end{equation}
with $a_i=1,...,\Nc$ if parton $i$ is a quark or an anti-quark and
$a_i=1,...,\Nc^2-1$ if parton $i$ is a gluon.

Clearly, in any QCD calculation, the color amplitudes, 
$\Col_1$ and $ \Col_2$, may be kept as they are, with color structure 
read off from the contributing Feynman diagrams. 
Alternatively -- and this is likely to be the preferred solutions
for more than a few partons -- they may be decomposed 
into a color basis (spanning set), such as a trace basis 
\cite{Paton:1969je,Dittner:1972hm,Cvi76,Cvitanovic:1980bu,
Berends:1987cv,Mangano:1987xk,Mangano:1988kk,Nagy:2007ty,Sjodahl:2009wx},
a color flow basis \cite{Maltoni:2002mq}
or a multiplet basis \cite{Keppeler:2012ih, DecompositionPaper}.

\section{Trace bases}
\label{sec:trace_bases}

One way of organizing calculations in color space is to use 
trace bases 
\cite{Paton:1969je,Dittner:1972hm,Cvi76,Cvitanovic:1980bu,
Berends:1987cv,Mangano:1987xk,Mangano:1988kk,Nagy:2007ty,Sjodahl:2009wx}.
To see that this is always possible, we note that 
the triple-gluon vertex can be expressed as 
\begin{eqnarray}
  \label{eq:f}
  i f^{abc}&=&\parbox{2cm}{\epsfig{file=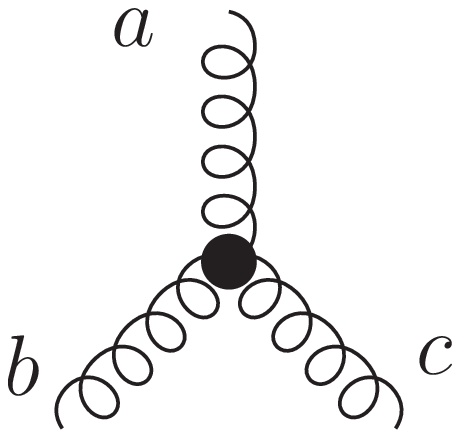,width=2cm}} =
  \frac{1}{\TR} \left[ 
    \parbox{2cm}{\epsfig{file=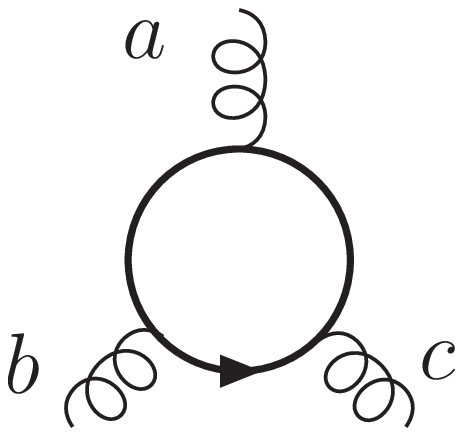,width=2cm}}
    - \parbox{2cm}{\epsfig{file=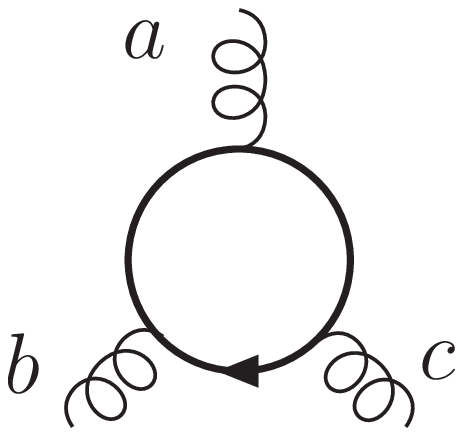,width=2cm}} \right] \\
  &=&\frac{1}{\TR}\left[ (t^a)^i{}_j (t^b)^j{}_k (t^c)^k{}_i -(t^b)^i{}_j (t^a)^j{}_k (t^c)^k{}_i  \right] 
  =\frac{1}{\TR}\left[ \tr[t^a t^b t^c] -\tr[t^b t^a t^c ] \right].\nonumber
\end{eqnarray}
where $\TR$ is the normalization of the SU($\Nc$) generators, 
$\tr(t^a t^b)=\TR\delta^{ab}$, commonly taken to be $1/2$ or $1$.

Using this relation on every triple-gluon vertex in any amplitude results  
in general in a sum of products of (connected) traces over SU($\Nc$) generators
and open quark-lines. More specifically, there is one open quark-line 
for every incoming quark/outgoing anti-quark and outgoing quark/incoming anti-quark.
(Note that, from a color perspective outgoing anti-quarks are equivalent to incoming 
quarks; we will here refer to them collectively as quarks. Similarly 
outgoing quarks are equivalent to incoming anti-quarks, and will be referred to as
anti-quarks.)

To further simplify the color structure, we may contract every internal
gluon propagator (which after application of \eqref{eq:f} connects quark-lines) using the Fierz 
(completeness) relation
\begin{eqnarray}
  \parbox{2cm}{\epsfig{file=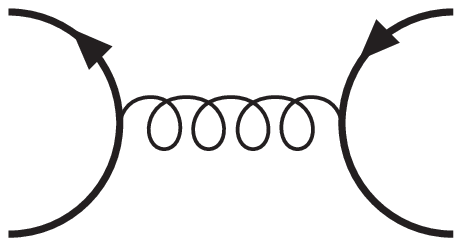,width=2cm}}
  &=&\TR \left[ 
    \parbox{1.5cm}{\epsfig{file=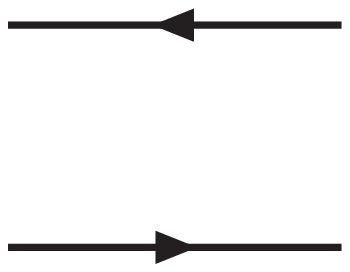,width=1.5cm}}
    - \frac{1}{\Nc} 
      \parbox{1.5cm}{\epsfig{file=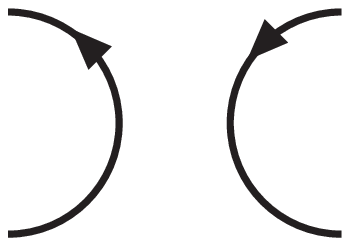,width=1.5cm}} \right].
\label{eq:Fierz}
\end{eqnarray}

From this we see that every amplitude in QCD, at any order, may be expressed
as a sum of products of open and closed quark-lines, where {\it all} gluon indices
are external indices. The set of all such products of quark-lines
can thus be used as a spanning set for any QCD process, for example,
for one $\qqbar$-pair and seven gluons, we may have color structures of
form 
\begin{eqnarray}
  \parbox{5.5 cm}{\epsfig{file=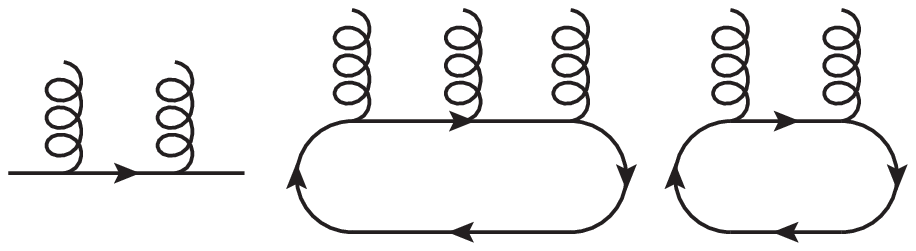,width=5.5cm}}
\end{eqnarray}
for all possible gluon permutations, as well as color structures with 
more or fewer traces.
We here refer to this type of basis as a {\it trace basis}, although we remark that 
when the number of gluons, $\Ng$, plus the number of
$\qqbar$-pairs, $\Nq$, exceeds $\Nc$, this spanning set is overcomplete,
and hence strictly speaking not a basis.
For $\Ng+\Nq\leq\Nc$, the bases are not overcomplete, but they are still
non-orthogonal, having non-diagonal scalar products being suppressed by powers of $\Nc$. 
Only in the $\Nc\to \infty$ limit, are these bases minimal and orthogonal.

As a simple, but non-trivial example, we may consider the basis needed for
$q_1 \qbar_2 \to g_3 g_4$. The basis, which is constructed by connecting quarks 
and gluons in all allowed ways \cite{Sjodahl:2009wx}
is given by\footnote{To enhance the similarity with C++, the vector numbering starts at 0 here.}
\begin{eqnarray}
  \label{eq:1q1g}
  \Vector^0_{q_1,q_2,g_3,g_4}&=& (t^{g_3} t^{g_4})^{q_1}{}_{q_2} 
  = \raisebox{0.4\height}{\parbox{3.5cm}{\epsfig{file=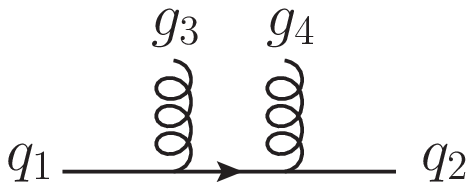,width=3.5cm}} }\nonumber \\
  \Vector^1_{q_1,q_2,g_3,g_4}&=&(t^{g_4} t^{g_3})^{q_1}{}_{q_2} 
  = \raisebox{0.4\height}{\parbox{3.5cm}{\epsfig{file=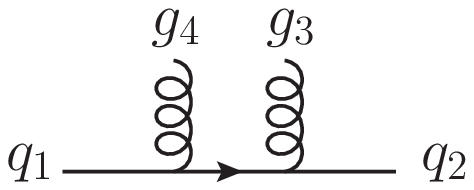,width=3.5cm}} }\\
  \Vector^2_{q_1,q_2,g_3,g_4}&=&\delta^{q_1}{}_{q_2}\tr(t^{g_3} t^{g_4})
  = \raisebox{0.4\height}{\parbox{3.9cm}{\epsfig{file=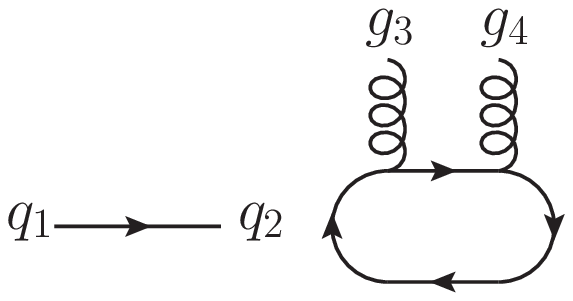,width=3.8cm}} }.\nonumber
\end{eqnarray}
To all orders in perturbation theory 
in the $\Nc\to \infty$ limit, one can prove that the number of 
basis vectors 
can be found
using the recursion relation \cite{Keppeler:2012ih}
\begin{equation}
  N_{\mbox{\tiny vec}}[\Nq, \Ng]
  = N_{\mbox{\tiny vec}}[\Nq, \Ng-1](\Ng-1+\Nq)
  + N_{\mbox{\tiny vec}}[\Nq, \Ng-2](\Ng-1) \, , 
  \label{eq:Nvqg}
\end{equation}
where
\begin{eqnarray}
  N_{\mbox{\tiny vec}}[\Nq, 0]&=&\Nq!\;,\;\;\;  
  N_{\mbox{\tiny vec}}[\Nq, 1]=\Nq \Nq!\;.
  \label{eq:start_cond}
\end{eqnarray}
In the gluon-only case, at tree-level, the only color structures
that can appear are traces over generators, meaning that a 
general tree-level gluon amplitude can be decomposed as 
\bea
\mathcal{M}(g_1,g_2,\dots, g_n)
=\hspace*{-0.2 cm}\sum_{\sigma \in S_{\Ng-1}}\tr(t^{g_1}t^{g_{\sigma_2}}\dots t^{g_{\sigma_n}})A(\sigma)
=\hspace*{-0.2 cm}\sum_{\sigma \in S_{\Ng-1}}\parbox{3.3cm}{\epsfig{file=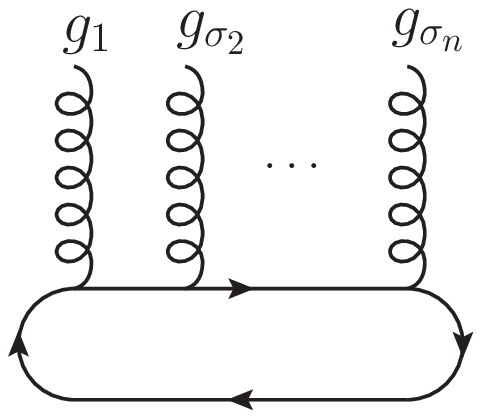,width=3.3 cm}}
\hspace*{-0.5 cm}A(\sigma).
\label{eq:Trace-decomposition}
\eea
That only fully connected color structures enter in tree-level gluon 
amplitudes can easily be understood from the decomposition of Feynman diagrams 
into basis vectors; upon application of \eqref{eq:f} all external
gluons remain attached to a quark-line, and -- while contracting internal 
gluons using the Fierz identity, \eqref{eq:Fierz} -- they remain connected to the same 
quark-line, as the color suppressed terms cancel each other out.
(This can be proved by a short calculation.)
The same cancellation appears for gluon exchange between a quark and a 
gluon, meaning that also tree-level color structures for one 
$\qqbar$-pair and $\Ng$ gluons must be of the ``fully connected'' form
of a trace that has been cut open, an open quark-line,
\bea
\mathcal{M}(q_1,g_3,\dots, g_n,\qbar_2)
= \hspace*{-0.2 cm}\sum_{\sigma\in S_{\Ng}}(t^{g_{\sigma_1}}t^{g_{\sigma_2}}\dots t^{g_{\sigma_n}})^{q_1}{}_{q_2}A(\sigma)
=\hspace*{-0.2 cm}\sum_{\sigma\in S_{\Ng}}\hspace*{-0.13 cm}\parbox{5.2cm}{\epsfig{file=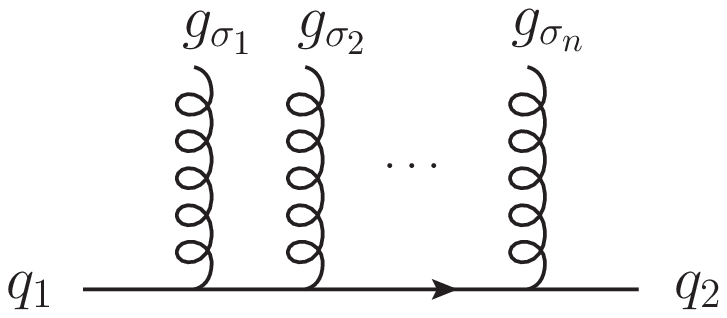,width=5.2 cm}}\hspace*{-0.8 cm}A(\sigma),\hspace*{-0.5 cm}
\nonumber\\
\label{eq:Trace-decomposition}
\eea
i.e., only the first two basis vectors in \eqref{eq:1q1g} are needed.
However, when the Fierz identity is applied directly 
to a gluon exchange between quarks, as in \eqref{eq:Fierz}, 
both terms do appear, and color structures with up to
$\Nq$ disconnected quark-lines may appear even at tree-level.

Starting from a trace basis tree-level color structure, for example a single 
trace over gluons, and exchanging a gluon between two partons 
may split off a disconnected color structure, such as in
\begin{equation}
\parbox{3.5cm}{\epsfig{file=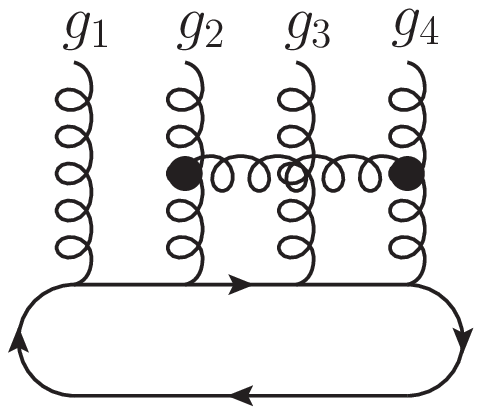,width=3.5cm}}
  = -\TR \parbox{3.7cm}{\epsfig{file=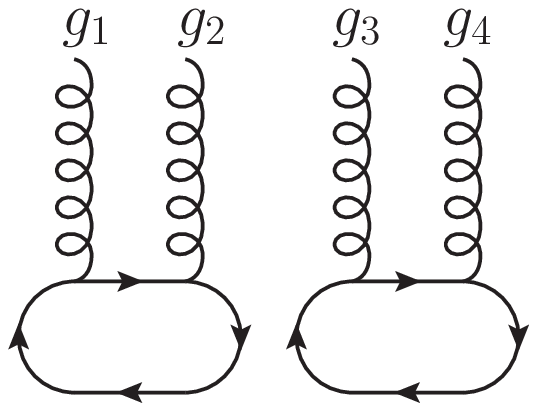,width=3.7cm}} \,
    -\TR \parbox{3.7cm}{\epsfig{file=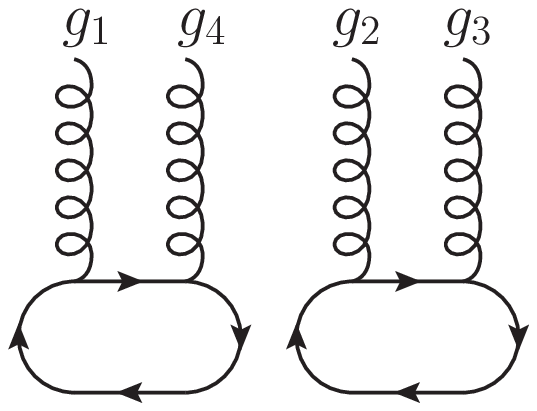,width=3.7cm}} \, .
\label{eq:GluonExchangeOnTrace}
\end{equation} 
Thus, counting to $l_g$ additional gluon exchanges
(on top of a tree-level diagram),
the color structures can not consist of more than
$\mbox{max}(1,\Nq)+l_g$ open and closed quark-lines, two in the above case.
In general, when any Feynman diagram is decomposed into a 
trace basis, there can be at most $\Nq+\lfloor \Ng/2 \rfloor$
quark-lines, since all gluons may be disconnected from the quarks,
but no gluon can stand alone in a trace, giving the factor $\lfloor \Ng/2 \rfloor$.

For NLO color structures having a quark-loop in the Feynman diagram, 
the quark-loop is necessarily connected to the remaining color structure
via at least one gluon exchange, unless the Feynman diagram consists 
of only gluons attached to one quark-loop, in which case the
trace basis decomposition is trivial. In general it may be connected
to more tree-level color structures, for example of form
\begin{equation}
  \parbox{5cm}{\epsfig{file=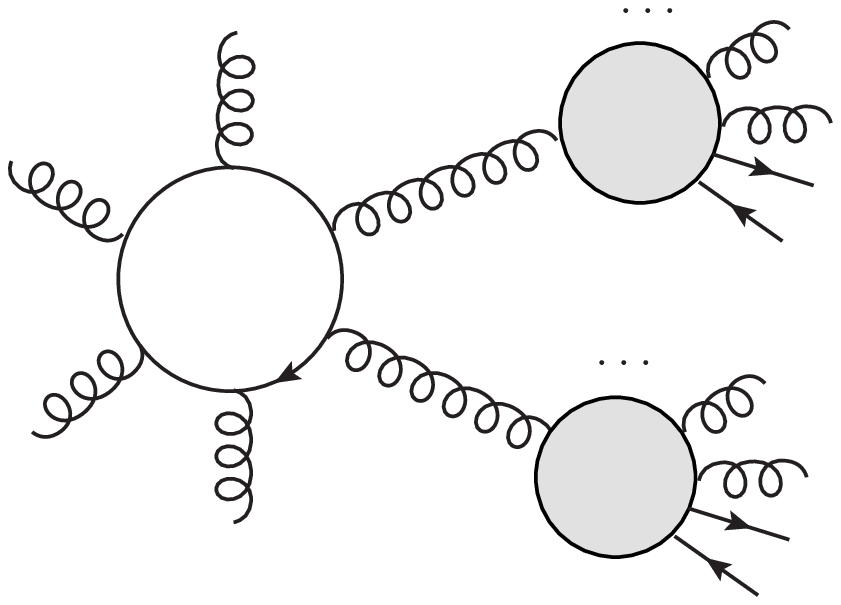,width=5cm}},
\end{equation} 
where the gray blobs denote any tree-level color structures. 
For each color structure connected 
to the quark-loop (gray blob), contracting the connecting gluon gives a 
term where the quark-loop is disconnected if the connecting gluon 
goes to a quark, the second term in
\begin{equation}
  \parbox{4.5cm}{\epsfig{file=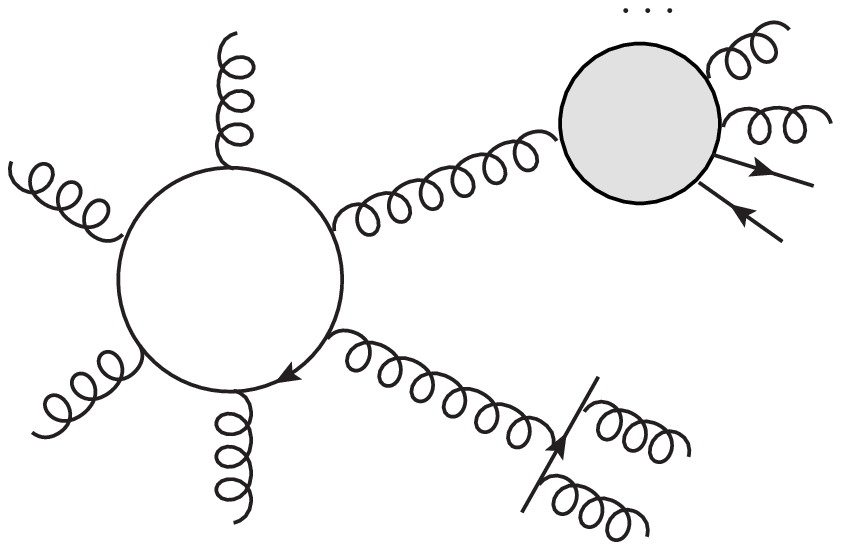,width=4.5cm}}
  =\TR\parbox{4.5cm}{\epsfig{file=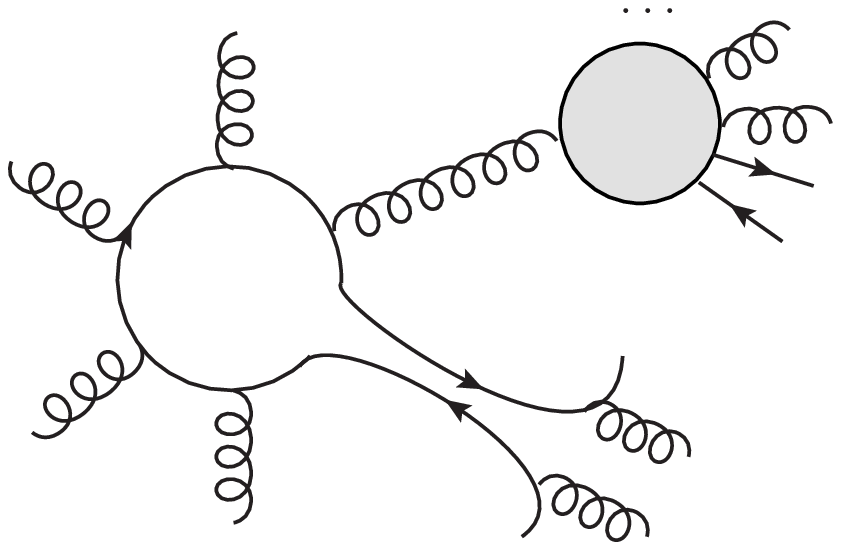,width=4.5cm}}
  -\frac{\TR}{\Nc}\parbox{4.5cm}{\epsfig{file=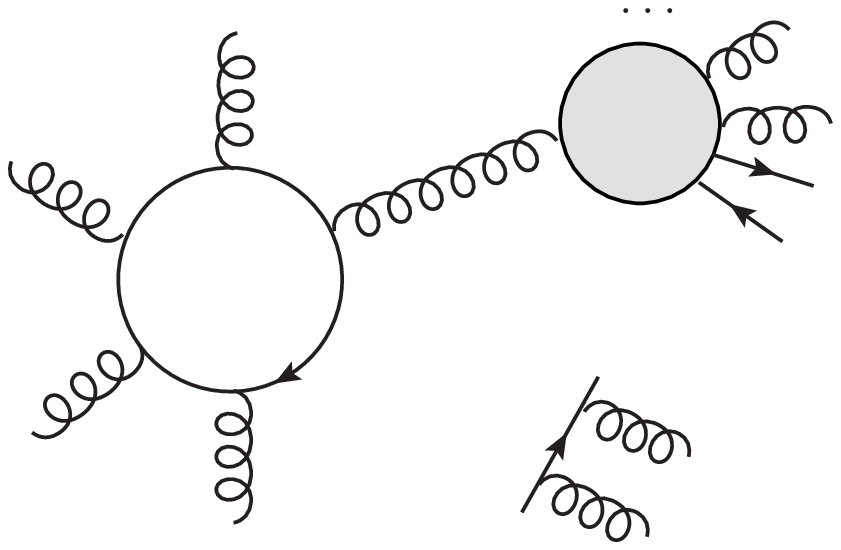,width=4.5cm}}.
\end{equation} 
If the quark-loop is not connected to any other color structure 
(there is no remaining gray blob in the equation above),
the resulting color structure thus contains one more 
single trace meaning that, decomposed in the trace basis, we 
get one more quark-line than at tree-level. 
(For purely gluonic processes, no additional trace will
appear.) 
The same argument can be applied repeatedly if the Feynman diagram contains 
more than one closed quark-loop, giving (similar to the gluon case) 
no more than $\mbox{max}(\Nq,1)+l_q$  quark-lines for $l_q$ quark-loops.

Finally, we note that, if, on top of the quark-loops,
there are also $l_g$ loops from gluon exchanges in the Feynman 
diagram, these may contribute with up to one additional quark-line each 
(such as in \eqref{eq:GluonExchangeOnTrace}), completing the argument that for  
$n_l=l_q+l_g$ loops, a general color structure in pure QCD can be written in 
terms of sums of products of closed and open quark-lines, where each
product contains at most  
$\mbox{min}[\mbox{max}(1,\Nq) +n_l, \Nq+\lfloor \Ng/2 \rfloor ]$ 
traces and open quark-lines.

\section{Computational strategy}
\label{sec:strategy}

The \ColorFull\, strategy for treating color space is based on the 
above observations, i.e.,

\begin{itemize}
\item For given external particles (quarks, anti-quarks and gluons), 
  we may always decompose the color space into a linear combination
  of closed and open quark-lines, as described in \secref{sec:trace_bases}.
  (Other color bases may be expressed in terms of linear combinations
  of these terms.)

\item We can always evaluate scalar products by first replacing 
  triple-gluon vertices using \eqref{eq:f}, then removing all gluon 
  propagators using \eqref{eq:Fierz}, and finally contracting the 
  remaining product of $\qqbar$-delta functions.
\end{itemize}
The above outlined procedure for calculating scalar products has the 
advantages of being conceptually easy, and of covering all 
contractions occurring in QCD.
It has the disadvantage of potentially giving a very large number of 
terms, since every replacement of a structure using \eqref{eq:f} 
may double the number of terms, and similarly, so does naive 
application of the Fierz identity, \eqref{eq:Fierz}.

In order to mitigate this potential explosion of terms upon index 
contraction, the \ColorFull\, 
procedure for scalar product contraction is:

\begin{enumerate}
\item Contract all quark-ends, giving a sum of products of {\it closed} quark-lines.

\item On the individual quark-lines, contract all neighboring gluons
  (giving a factor $\CF$) and all next to neighboring gluons, giving a factor
  $-\TR/\Nc$ each, as only the color suppressed term in the Fierz identity,
  \eqref{eq:Fierz}, survives. Also, contract traces of two gluons,
  $\tr(t^{g_1}t^{g_2}) = \delta^{g_1 g_2}/\TR$.

\item Look for gluons to contract within the quark-lines.
  Each such contraction may give rise to two terms, 
  but at least the new traces tend to be shorter.

\item Look for gluon index contractions between the quark-lines.
\end{enumerate}
In this way all color indices can be iteratively contracted.
To further speed up calculations, \ColorFull\, can use memoization 
to save calculated topologies, meaning that contractions that
differ only by a relabeling of indices are performed only once.
This significantly speeds up the calculations.

\section{ColorFull at a glance}
\label{sec:colorfull}

\ColorFull\, is designed to handle the contraction of 
QCD color indices, to decompose the QCD color space using (trace) 
basis vectors, and to describe the effect of gluon 
exchange and gluon emission. In particular, \ColorFull\, can,
for arbitrary $\Nc$:

\begin{itemize}
\item Square any QCD amplitude and calculate any
  interference term.

\item Create a trace basis for any number of quarks and gluons,
and to arbitrary order.

\item Read in and write out color bases, including non-trace bases.

\item Calculate scalar product matrices, i.e., the matrices of scalar products
  between the basis vectors, write these out and read them in again.

\item  Describe the effect of gluon exchange, including calculating the color
  soft anomalous dimension matrices.

\item  Describe the effect of gluon emission.

\end{itemize}
\ColorFull\, can also be interfaced to \Herwig\, ($\geq$ 2.7) 
via \Matchbox\, \cite{Platzer:2011bc} and can thus easily be used for event generation 
with \Herwig.

\subsection{Technical overview}
\label{sec:overview}

\begin{table}[h]
 \caption{\label{tab:classes} 
    Below is a complete list of the \ColorFull\, classes, ordered 
    by dependence, such that, a certain class only depends on classes
    standing above it in the list.
  }
  \begin{tabular}[t]{ | p{4.5cm} |  p{10.9cm} |}
    \hline \hline
    Class & Functionality
    \\\hline \hline
    \Monomial & Class for containing a single term of form
    $\Nctt^\powNc \times \TRtt^\powTR \times \CFtt^\powCF \times \texttt{int\_part} \times \texttt{\cnumpart}$
    and associated functions.
    \\\hline

    \Polynomial & Class for containing a sum of \Monomials\,
    and associated functions.
    \\\hline

    \Polyvec & Class for containing a vector of \Polynomials\,
    and associated functions.
    \\\hline

    \Polymatr & Class for containing a matrix of \Polynomials\,
    and associated functions.
    \\\hline

    \Quarkline & Class for containing a single closed or open quark-line
    multiplying a \Polynomial\, and associated functions.
    For example a term of the form $\tr[t^{g_1}t^{g_2}t^{g_3}t^{g_4}]$
    $\sim \texttt{(1,2,3,4)}$
    or of the form $(t^{g_3}t^{g_4})^{q_1}_{q_2}\sim \texttt{\{1,3,4,2\}}$ times a \Polynomial.
    \\\hline

    \Colstr & Class for containing a product of \Quarklines\, multiplying a \Polynomial,
    and associated functions.
    For example a polynomial $(\Nc^2-1)$ may multiply the \Quarklines, 
    $\delta^{q_1}{}_{q_2}\delta^{g_1 g_4}$, in total giving the \Colstr\,
    $\sim$ \texttt{(\Nctt$^{}$2-1) [\{1,2\}(3,4)]} 
    \\\hline

    \Colamp & Class for containing a sum of \Colstrs\,
    such as
    $\tr[t^{g_1}t^{g_2}t^{g_3}t^{g_4}]+\tr[t^{g_1}t^{g_2}]\tr[t^{g_3}t^{g_4}]$
    $\sim$\texttt{(1,2,3,4)+(1,2)(3,4)}
    and associated functions.
    \\\hline

    \Colfunctions & Library class containing functions for 
    performing numerical evaluations,
    taking the leading $\Nc$ limit, evaluating scalar products, 
    and describing gluon emission and exchange.
    \\\hline

    \Colbasis & Base class for all basis classes, see below.
    \\\hline

    \Tracetypebasis & Base class for \Tracebasis\, and
    \Treelevelgluonbasis.
    \\\hline

    \Tracebasis & Class for creating and using trace bases.
    Perhaps the most important basis class.
    \\\hline

    \Treelevelgluonbasis & Class for bases needed for tree-level gluon calculations,
    where each basis vector is a single trace plus an implicit complex conjugate.
    \\\hline

    \Orthogonalbasis & Class for utilizing the benefits of orthogonal bases
    such as multiplet bases \cite{Keppeler:2012ih, DecompositionPaper}.
    \\\hline

  \end{tabular}
\end{table}

This section is intended to give an overview of \ColorFull. 
For examples, the reader is referred to \secref{sec:standalone}, for 
overviews of the classes to \secref{sec:code}, 
and for details, the tables in the \appref{sec:member_functions}, as 
well as the online Doxygen documentation.

As contraction of indices gives rise to polynomials
in $\Nc$, $\TR$ and $\CF$, \ColorFull\, by necessity needs minimal classes for 
dealing with such polynomials. This is implemented in the classes
\Monomial\, (a single term in a polynomial\footnote{Throughout the \ColorFull\, 
documentation terms of form constant$\times \Nc^a\, \TR^b \,\CF^c$ will 
be referred to as monomials, despite the possible occurrence 
of negative powers. Similarly sums of such terms will be referred to 
as polynomials.}), \Polynomial\,
(a sum of \Monomials), \Polyvec\, (a vector of \Polynomials)
and \Polymatr\, (a matrix of \Polynomials).

For the color structure itself, \ColorFull\, uses the class 
\Quarkline\, for treating an individual closed or open quark-line,
a class \Colstr\, for treating a product of \Quarklines\,
and a class \Colamp\, for treating a general color amplitudes,
i.e., a linear combination of \Colstrs.

For performing numerical evaluations, taking the leading $\Nc$
limit, evaluating scalar products, and describing gluon 
emission and exchange, \ColorFull\, has a library class 
\Colfunctions.

Finally, \ColorFull\, has classes for describing color bases.
The classes intended for the user, 
\Tracebasis, \Treelevelgluonbasis\, and \Orthogonalbasis, are derived 
from the base class \Colbasis\, (in the case of the first two
via the class \Tracetypebasis).

All \ColorFull\, classes are listed in \tabref{tab:classes} 
according to dependencies, meaning that each class only depends
on classes listed above it.
The next section will give an introduction to using \ColorFull\,
in stand-alone mode.

\section{Stand-alone usage}
\label{sec:standalone}

\ColorFull\, is mainly designed to deal with relatively large color 
spaces where it is advantageous to use bases in which color 
structures coming from Feynman diagrams -- or alternative recursive
strategies -- are decomposed.

For trace bases, the bases may be automatically created 
by \ColorFull.
For example, a basis for \texttt{1} $\qqbar$-pair, \texttt{3} 
gluons and \texttt{0} loops (in pure QCD) can be created using
\begin{eqnarray}
  & &  \texttt{Trace\_basis MyBasis(1,3,0);}.
\end{eqnarray}
The last argument is provided to avoid carrying around basis 
vectors for which the kinematic factors vanish to a certain
order in perturbation theory. It can be skipped upon which
an all order basis vector is created.

To view the resulting basis, it can be written out to a stream
(\texttt{cout}) or to a file, either with default filename 
(no argument) or a user supplied name,
\begin{eqnarray}
  & &  \texttt{cout << MyBasis;}\nonumber \\
  & &  \texttt{MyBasis.write\_out\_Col\_basis("ColorResults/MyBasis");}
\end{eqnarray}
resulting in
\begin{eqnarray}
  &\texttt{0}\;\;\;\; & \texttt{[\{1,3,4,5,2\}]} \nonumber \\
  &\texttt{1}\;\;\;\; & \texttt{[\{1,3,5,4,2\}]} \nonumber \\
  &\texttt{2}\;\;\;\; & \texttt{[\{1,4,3,5,2\}]} \nonumber \\
  &\texttt{3}\;\;\;\; & \texttt{[\{1,4,5,3,2\}]} \nonumber \\
  &\texttt{4}\;\;\;\; & \texttt{[\{1,5,3,4,2\}]} \nonumber \\
  &\texttt{5}\;\;\;\; & \texttt{[\{1,5,4,3,2\}]}
\end{eqnarray}
where, for example,
\begin{equation}
\texttt{[\{1,3,4,5,2\}]}=(t^{g3}t^{g4}t^{g5})^{q1}{}_{q2}= 
\parbox{3cm}{\epsfig{file=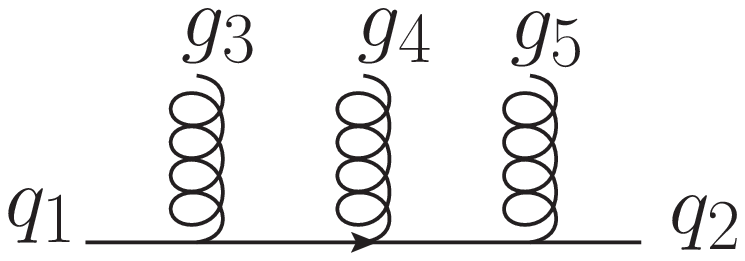,width=3cm}}.
\end{equation}
If no argument is supplied to \texttt{MyBasis.write\_out\_Col\_basis()},
the basis is written out to a file with a default filename, in this
case \texttt{CF\_TB\_q\_1\_g\_3}. 
The bases can thus be written out and saved for future purposes.
For reading in a basis 
\texttt{MyBasis.read\_in\_Col\_basis("path/to/filename")}
can be used. 

The option of reading in bases is particularly useful for the 
\Orthogonalbasis\, class. Presently \ColorFull\, can not automatically
create orthogonal multiplet bases, such as in \cite{Keppeler:2012ih},
but externally created (orthogonal) bases, with basis vectors
expressed in terms of sums of products of traces, may be read
in and used. For the \Orthogonalbasis\, class, the orthogonality
is then utilized to significantly speed up the the calculation of 
scalar products.
For a non-orthogonal basis it is necessary to evaluate 
all scalar products between all basis vectors, which can be
done as
\bea 
\texttt{MyBasis.scalar\_product\_matrix();}.
\eea
\Polynomial\, and \texttt{double} versions of the scalar product 
matrix are then calculated and saved in the member variables 
\texttt{P\_spm} and \texttt{d\_spm}.
For larger bases, the evaluation can be sped up by only 
calculating the numerical version using 
\texttt{MyBasis.scalar\_product\_matrix\_num()}.
The content of \texttt{P\_spm} and \texttt{d\_spm} can be 
explored by using the \texttt{<<} operator, but it may also be 
saved to a file using the \Colbasis\, member functions 
\texttt{write\_out\_P\_spm()} and \texttt{write\_out\_d\_spm()}
which write out the result in \ColorFull\, and Mathematica, readable format,
by default to a file with a default filename in the directory \texttt{ColorResults}.
Alternatively the user may supply a filename as argument.

While the intended usage of \ColorFull\, is to use color bases,
\ColorFull\, can also directly define color amplitudes 
\begin{eqnarray}
  & & \texttt{Col\_amp Ca1("[\{1,3,2\}(4,5)]");}\quad\quad\; \slash\slash\;\;  (t^{g3})^{q1}{}_{q2} \tr(t^{g4}t^{g5} ) \nonumber \\
  & & \texttt{Col\_amp Ca2("[\{1,3,4,5,2\}]");} \quad\quad\quad \slash\slash\;\;  (t^{g3}t^{g4}t^{g5})^{q1}{}_{q2}
\end{eqnarray}
and evaluate scalar products using the \texttt{scalar\_product} 
member function in the \Colfunctions\, library class
\begin{eqnarray}
  & & \texttt{Col\_functions Col\_fun;}\nonumber\\
  & & \texttt{Col\_fun.scalar\_product(Ca1,Ca1);}\\
  & & \texttt{Col\_fun.scalar\_product(Ca1,Ca2);}\nonumber
\end{eqnarray}
giving the results 
\texttt{TR*Nc\^{}{2}*CF\^{}{2}} and 
\texttt{TR*Nc*CF\^{}{2}}, respectively.

In several contexts, such as parton showers, cancellation of infrared 
singularities in NLO calculations, and recursive methods \cite{Du:2015apa} for 
calculating amplitudes, it is of interest to know the effect of
gluon emission on a color structure. This can be calculated by using 
the \Colfunctions\, member function \texttt{emit\_gluon}
\begin{eqnarray}
  \texttt{Col\_fun.emit\_gluon(Ca1,3,6);}
\end{eqnarray}
resulting in
\bea
\texttt{[\{1,3,6,2\}(4,5)]-[\{1,6,3,2\}(4,5)]}.
\eea
The sign conventions here and elsewhere are such that every gluon inserted
after the emitter in the quark-line comes with a plus sign
and every gluon inserted before comes with a minus sign.
In a basis-decomposed calculation, one would be interested
in this result decomposed in the large basis required for 
one additional gluon.
Having a trace basis for this larger vector space, this decomposition
can be calculated using 
\begin{eqnarray}
  & &\texttt{Trace\_basis LargerBasis(nq,ng+1);}\\
  & &\texttt{LargerBasis.new\_vector\_numbers(Cs, emitter);}
\end{eqnarray}
where \texttt{Cs} is the \Colstr\, (basis vector in the smaller 
trace basis) from which the parton \texttt{emitter} emits a new gluon.

Similarly the effect of gluon exchange on a color structure 
is of interest. 
\ColorFull\, offers several functions for dealing with this.
Starting with an amplitude, the new amplitude after gluon 
exchange can be obtained as 
\begin{eqnarray}
& &\texttt{Col\_fun.exchange\_gluon(Ca1, 1, 4);}
\end{eqnarray}
resulting in \texttt{TR[\{1,4,5,3,2\}] - TR[\{1,5,4,3,2\}]}. 
\ColorFull\, also has a function for directly calculating the 
color correlator, i.e., given a color amplitude $\left|\Col\right>$, the
quantity
$\left<\Col| \mathbf{T}_i \cdot \mathbf{T}_j |\Col\right>$ where $\mathbf{T}_i$ describes the effect
of attaching a gluon to parton $i$. For example 
\begin{eqnarray}
  \texttt{Col\_fun.color\_correlator(Ca1, 1, 2);}
\end{eqnarray}
results in the 
\Polynomial\, \texttt{TR\^{}(2)*Nc*CF\^{}(2)}.

In some situations, such as soft gluon resummation, it is also 
useful to calculate the soft anomalous dimension matrix,
i.e., to have the result of gluon exchange 
on any basis vector decomposed into the basis.
This can be computed automatically using 
the \Colbasis\, member function \texttt{color\_gamma}.
The result is contained in a $\Polymatr$. 
For this we need the full basis since color structures
not present at LO will appear at NLO, etc. Coding
\begin{eqnarray}
  & &  \texttt{Trace\_basis MyFullBasis(1,3);}\nonumber \\
  & &  \texttt{MyFullBasis.color\_gamma(1,4);}
\end{eqnarray}
for gluon exchange between the partons \texttt{1} and \texttt{4} 
results in the matrix
\begin{eqnarray}
  \label{eq:gamma14}
  & &\texttt{\{\{                  0,                  0,                  0,                  0,                  0,                  0,                  0,                  0,            1*-1 TR,                  0,                   0\},}\nonumber \\
  & &\texttt{\{                  0,                  0,                  0,                  0,                  0,                  0,                  0,                  0,                  0,                  0,             1*-1 TR\},}\nonumber \\
  & &\texttt{\{                  0,                  0,            1 TR Nc,                  0,                  0,                  0,                  0,                  0,               1 TR,                  0,                1 TR\},}\nonumber \\
  & &\texttt{\{                  0,                  0,                  0,            1 TR Nc,                  0,                  0,               1 TR,                  0,                  0,               1 TR,                   0\},}\nonumber \\
  & &\texttt{\{                  0,                  0,                  0,                  0,                  0,                  0,                  0,                  0,                  0,            1*-1 TR,                   0\},}\nonumber \\
  & &\texttt{\{                  0,                  0,                  0,                  0,                  0,                  0,            1*-1 TR,                  0,                  0,                  0,                   0\},}\nonumber \\
  & &\texttt{\{                  0,                  0,                  0,                  0,                  0,            1*-1 TR,                  0,                  0,                  0,                  0,                   0\},}\nonumber \\
  & &\texttt{\{                  0,               1 TR,                  0,                  0,               1 TR,                  0,                  0,            1 TR Nc,                  0,                  0,                   0\},}\nonumber \\
  & &\texttt{\{            1*-1 TR,                  0,                  0,                  0,                  0,                  0,                  0,                  0,                  0,                  0,                   0\},}\nonumber \\
  & &\texttt{\{                  0,                  0,                  0,                  0,            1*-1 TR,                  0,                  0,                  0,                  0,                  0,                   0\},}\nonumber \\
  & &\texttt{\{                  0,            1*-1 TR,                  0,                  0,                  0,                  0,                  0,                  0,                  0,                  0,                   0\}\}}. 
\end{eqnarray}
In this way all the soft anomalous dimension matrices needed
in \cite{Gerwick:2014gya} can easily be recalculated.

While the number of colors in QCD is three, \ColorFull\,
can deal with any $\Nc$, both algebraically and numerically. 
Numerical evaluation is handled by the \Colfunctions\,
class, using the (private) member variables \Nctt, \TRtt\,
and \CFtt.
Thus also $\TR$ and $\CF$ can be changed independently.
The reason for keeping $\CF$ as a parameter technically
independent of $\Nc$ is that this allows for keeping
the color suppressed part of $\CF$, $-\TR/\Nc$, in a 
consistent way. This choice has proved useful for accounting 
for sub-leading $\Nc$ effects in several phenomenological 
studies \cite{Platzer:2012np, Platzer:2013fha}.

To numerically evaluate 
a \Monomial, \Polynomial, \Polyvec\, or \Polymatr, the
\Colfunctions\, member functions \texttt{double\_num}
are used, for example we may want the $\Nc=3$ version of a 
scalar product
\begin{eqnarray}
  \texttt{Col\_fun.double\_num(Col\_fun.scalar\_product(Ca1,Ca1)} );
\end{eqnarray}
giving \texttt{8}. 

For comparison, it is of interest to evaluate (squared)
amplitudes in the limit $\Nc \to \infty$. 
For taking the leading $\Nc$ limit of any of the polynomial 
classes \Polynomial, \Polyvec\, or \Polymatr,
the \Colfunctions\, member function(s) \texttt{leading}
can be used. For example we may take the leading $\Nc$
limit of the matrix in \eqref{eq:gamma14}, 
\begin{eqnarray}
  \texttt{Col\_fun.leading(MyFullBasis.color\_gamma(1,4));}
\end{eqnarray}
resulting in the diagonal matrix 
\begin{eqnarray}
  \label{eq:gamma14limit}
& &\texttt{\{\{                   0,                  0,                  0,                  0,                  0,                  0,                  0,                  0,                  0,                  0,                   0\},}\nonumber \\
& &\texttt{\{                   0,                  0,                  0,                  0,                  0,                  0,                  0,                  0,                  0,                  0,                   0\},}\nonumber \\
& &\texttt{\{                   0,                  0,            1 TR Nc,                  0,                  0,                  0,                  0,                  0,                  0,                  0,                   0\},}\nonumber \\
& &\texttt{\{                   0,                  0,                  0,            1 TR Nc,                  0,                  0,                  0,                  0,                  0,                  0,                   0\},}\nonumber \\
& &\texttt{\{                   0,                  0,                  0,                  0,                  0,                  0,                  0,                  0,                  0,                  0,                   0\},}\nonumber \\
& &\texttt{\{                   0,                  0,                  0,                  0,                  0,                  0,                  0,                  0,                  0,                  0,                   0\},}\nonumber \\
& &\texttt{\{                   0,                  0,                  0,                  0,                  0,                  0,                  0,                  0,                  0,                  0,                   0\},}\nonumber \\
& &\texttt{\{                   0,                  0,                  0,                  0,                  0,                  0,                  0,            1 TR Nc,                  0,                  0,                   0\},}\nonumber \\
& &\texttt{\{                   0,                  0,                  0,                  0,                  0,                  0,                  0,                  0,                  0,                  0,                   0\},}\nonumber \\
& &\texttt{\{                   0,                  0,                  0,                  0,                  0,                  0,                  0,                  0,                  0,                  0,                   0\},}\nonumber \\
& &\texttt{\{                   0,                  0,                  0,                  0,                  0,                  0,                  0,                  0,                  0,                  0,                   0\}\}}.
\end{eqnarray}
For an expression containing $\CF$, 
the color suppressed term $-\TR /\Nc$ can be kept
in numerical evaluation by setting \texttt{full\_CF}
to \texttt{true}, \texttt{Col\_fun.set\_full\_CF(true)}.

\section{Interfacing to Herwig via Matchbox}
\label{sec:matchbox}

\ColorFull\, can also be interfaced to \Herwig\, ($\geq $2.7) \cite{Bellm:2013lba}
via the \Matchbox\, component \cite{Platzer:2011bc}, and 
can be used to treat the hard interaction, as in for 
example \cite{Campanario:2013fsa}, as well as the 
parton shower itself \cite{Platzer:2012np}.
When linked to \Herwig, \ColorFull\, is hooked into the 
\texttt{boost} linear algebra package, enabling
a very efficient treatment of numerical linear algebra.
From the next major release of \Herwig,
\ColorFull\, will be directly shipped with the \Herwig,
code.

\section{ColorFull code}
\label{sec:code}

\subsection{Operators}
\label{sec:operators}
In order to simplify calculations and increase readability, \ColorFull\,
defines a standard set of operators, listed in \tabref{tab:operators}.
This includes the arithmetic operators \texttt{+}, \texttt{-} and \texttt{*}
as well as the comparison operators, \texttt{==} and \texttt{!=}, 
and the stream operator \texttt{<<} .
The comparison operators, \texttt{==} and \texttt{!=}, work by comparing
data entries term by term, both for the polynomial classes
and for the color carrying classes. Thus, for example, two
\Polynomials\, only differing by the order of terms are not
considered equal.

\begin{table}[t]
  \caption{\label{tab:operators} 
    Standard operators for color structure and polynomial classes.
    (Operators for both orders of arguments are defined for the 
    arithmetic operators \texttt{+}, \texttt{-} and \texttt{*}.)}
  \begin{tabular}[t]{ | p{1.5cm} |  p{13.9cm} |}
    \hline \hline
    Operator & Classes \\ 
    \hline\hline
    \texttt{+} and \texttt{-} & 
    \begin{tabular}{l}
      \Polynomial\, $\texttt{+/-}$ (\Monomial, \Polynomial)\\
      \Colamp\, $\texttt{+/-}$ (\Colstr, \Colamp)
    \end{tabular}
    \\\hline

    \texttt{+=} & 
    \begin{tabular}{l}
      \Polynomial\, $\texttt{+=}$ (\Monomial, \Polynomial)\\
      \Colamp\, $\texttt{+=}$ (\Colstr, \Colamp)
    \end{tabular}
    \\\hline

    \texttt{*} &
    \begin{tabular}{l}
      \Monomial\texttt{*}(\integer, \cnum, \double, \Monomial)\\
      \Polynomial\texttt{*}(\integer, \cnum, \double, \Monomial, \Polynomial)\\
      \Quarkline\texttt{*}(\integer, \cnum, \double, \Monomial, \Polynomial, \Quarkline)\\
      \Colstr\texttt{*}(\integer, \cnum, \double, \Monomial, \Polynomial, \Quarkline, \Colstr)\\
      \Colamp\texttt{*}(\integer, \cnum, \double, \Monomial, \Polynomial, \Quarkline, \Colstr,\\
      \Colamp)
    \end{tabular}
    \\\hline
    
    \texttt{*=} &
    \begin{tabular}{l}
      \Monomial\texttt{*=}(\integer, \cnum, \double, \Monomial)\\
      \Polynomial\texttt{*=}(\integer, \cnum, \double, \Monomial, \Polynomial)\\
      \Colamp\texttt{*=}(\Colstr, \Colamp)\\
    \end{tabular}
    \\\hline

    \texttt{==} &
    \begin{tabular}{l}
      \Monomial\, \texttt{==} \Monomial\\
      \Polynomial\, \texttt{==} \Polynomial\\
      \Polyvec\, \texttt{==} \Polyvec\\
      \Polymatr\, \texttt{==} \Polymatr\\
      \Quarkline\, \texttt{==} \Quarkline\\
      \Colstr\, \texttt{==} \Colstr\\
      \Colamp\, \texttt{==} \Colamp\\
    \end{tabular}
    \\\hline

    \texttt{!=} &
    \begin{tabular}{l}
      \Monomial \texttt{!=} \Monomial\\
      \Polynomial \texttt{!=} \Polynomial\\
      \Polyvec\, \texttt{!=} \Polyvec\\
      \Polymatr\, \texttt{!=} \Polymatr\\
      \Quarkline \texttt{!=} \Quarkline\\
      \Colstr \texttt{!=} \Colstr\\
      \Colamp \texttt{!=} \Colamp\\
    \end{tabular}
     \\\hline

     $\texttt{<}$ &
     \begin{tabular}{l}
       \Monomial\, $\texttt{<}$ \Monomial \\
       The \Monomials\, are ordered first according to $\powNc+\powCF$, \\
       then according to $\powNc$ (for same $\powNc+\powCF$), then according to \\
       \intpart*\texttt{abs}(\cnumpart), then according to \intpart\, and finally according to\\
       $\powTR$. This is thus {\it not} a magnitude operator.
     \end{tabular}
    \\\hline

    $\texttt{<<}$ &
    \begin{tabular}{l}
       \Monomial,
	\Polynomial,
	\Polyvec,
	\Polymatr,
	\Quarkline,
	\Colstr,
	\Colamp,\\
	\Colbasis
     \end{tabular}
    \\\hline
  \end{tabular}
\end{table}

\subsection{Classes for Polynomials}
\label{sec:polynomialclasses}

The result of color index contraction can always be written as a 
sum of terms of the form $\Nc{}^a \TR{}^b \CF{}^c  \,\times \mbox{constant}$
where we allow for negative integers $a,b,c$. 
For the purpose of treating contracted color structures, 
\ColorFull\, has a minimalistic set of classes for 
basic manipulation of color factors arising when contracting
color indices. 
One such term is defined as a \Monomial, and a sum of such terms as 
a \Polynomial. To decompose vectors in a color space we also need 
a vector of \Polynomials, contained in a \Polyvec, and for a 
scalar product matrix (or soft anomalous dimension matrix) we need
a matrix of polynomials, a \Polymatr.

Apart from the \Monomial\, class, these classes 
have the actual polynomial information contained in uncapitalized
typedefs carrying the same name as the class in question. There
are thus \polynomial, \polyvec, and \polymatr\, typedefs.
For manipulating the polynomial classes, the operators listed in
\tabref{tab:operators} may be used. For example, we may 
multiply a \Polynomial\, and a \Monomial.

\subsubsection{Monomial}
\label{sec:monomial}

The simplest class for containing contracted color information is 
a \Monomial
\begin{equation}
\Monomial \sim  \Nctt^\powNc \, \TRtt^\powTR \, \CFtt^\powCF \times \intpart \times \cnumpart.
\end{equation}
The exponents $\powNc$, $\powTR$ and $\powCF$, as well as
\intpart\, are of type \texttt{int} and \cnumpart\, is a \texttt{cnum}, 
an \texttt{std::complex<double>}. These member variables 
carry the information about the actual monomial.

By default, using the standard constructor, \Monomial\texttt{()},
a \Monomial\, is set to 1, by letting 
\texttt{\cnumpart=1}, \texttt{\intpart=1} and setting all powers 
to 0\footnote{A default \Monomial\, is thus the neutral element 
under multiplication, rather than under addition.}.
A \Monomial\, with different integer part can be obtained by 
using the constructor taking an \texttt{int} or as argument,
and a general \Monomial\, can be constructed using the 
string constructor (see \tabref{tab:monomial} in \appref{sec:member_functions}).

\subsubsection{Polynomial}
\label{sec:polynomial}

A sum of \Monomials\, is contained in the class \Polynomial
\begin{eqnarray}
  \begin{tabular}{ccccc}
    \Polynomial & $\sim$ &  $\Monomial_0$ +$\Monomial_1$+ ...\\
    & &\texttt{poly} &      
  \end{tabular}
\end{eqnarray}
where, technically, the polynomial information is stored
in the public member \poly, of type 
\polynomial=\texttt{std::vector<Monomial>}.
\ColorFull\, is not designed for manipulating polynomials
but \Polynomial\, nevertheless has a minimalistic 
\texttt{simplify()} function which collects terms with 
equal power of $\Nc$, $\TR$ and $\CF$. It also has 
a \texttt{remove\_CF()} function for replacing 
$\CF$ with $\TR \Nc-\TR/\Nc$.
Details about the \Polynomial\, class can be found in
\tabref{tab:polynomial} in \appref{sec:member_functions}.

\subsubsection{Poly\_vec}
\label{sec:polyvec}

For dealing with vectors of \Polynomials, \ColorFull\, has a class \Polyvec
\begin{eqnarray}
  \begin{tabular}{ccccc}
    \Polyvec & $\sim$ &  ($\Polynomial_0$, $\Polynomial_1$, ...)\\
    & & \texttt{pv}       
  \end{tabular}.
\end{eqnarray}
\Polyvec\, is a container class for functions acting on vectors
of \Polynomials\, whereas the actual information is stored in the
\polyvec=\texttt{typedef std::vector<Polynomial>} data member,
\texttt{pv}.
The member functions of \Polyvec, which include \texttt{simplify()}
and \texttt{remove\_CF()}, can be found in \tabref{tab:polyvec} 
in \appref{sec:member_functions}.

\subsubsection{Poly\_matr}
\label{sec:polymatr}

Finally, there is a class for treating matrices of \Polynomials,
a \Polymatr, 
\begin{eqnarray}
  \begin{tabular}{ccc}
    \Polymatr & $\sim$ &  \{\{$\Polynomial_{00}$, $\Polynomial_{01}$, ..., $\Polynomial_{0n}$\}\\
               &       & \vdots\\
    &       &  \;\;\;\;\{$\Polynomial_{m0}$, $\Polynomial_{m1}$, ..., $\Polynomial_{mn}$\}\}\\
    &       &                          \texttt{pm}    
  \end{tabular}
\end{eqnarray}
which stores the matrix information as a 
vector of vectors of \Polynomials, a 
\texttt{\polymatr=typedef std::vector<Poly\_vec>}.
Again, details can be found in the appendix, 
in \tabref{tab:polymatr}.

\subsection{Classes containing the color structure}
\label{sec:colorclasses}

This section describes the building blocks used by \ColorFull\, to 
treat the color structure. {\it All} classes used to carry the 
color structure, \Quarkline, \Colstr, \Colamp,
including the basis classes \Colbasis\,  -- from which
\Tracetypebasis, \Tracebasis, \Treelevelgluonbasis\, and
\Orthogonalbasis\, are derived -- have the property that the actual 
information of the color structure is contained in a type with 
corresponding name, whereas the class acts as a container
for related functions. Thus, for example, the color 
amplitude information in a \Colamp\,
is contained in a \colamp\, variable.

\subsubsection{Quark\_line}
\label{sec:quarkline}

The most basic color carrying class is a \Quarkline.
Loosely speaking, a \Quarkline\, is a \quarkline\, 
times a \Polynomial,
\begin{equation}
  \begin{tabular}{ccccc}
    \Quarkline & $\sim$ & \Polynomial & $\times$  & \quarkline\,\\
    & & \texttt{Poly}       & $\times$  & \texttt{ql},\\
  \end{tabular}
\end{equation}
where the \quarkline\, (a \texttt{std::vector<int>}) \texttt{ql} contains the actual quark-line 
information, together with the 
boolean variable \texttt{open} which is \texttt{true} if the quark-line is
open, and \texttt{false} for a trace over gluon indices. The \quarkline\,
is multiplied by a \Polynomial.
For example, we may have the closed \Quarkline
\begin{eqnarray}
\texttt{(Nc\^{}2-1) (1,2,3,4)} \sim (\Nc^2-1) \; \tr(t^{g_1}t^{g_2}t^{g_3}t^{g_4})
\end{eqnarray}
or the open \Quarkline
\begin{eqnarray}
\texttt{Nc TR \{1,3,4,2\} }&\sim& \Nc \TR\; (t^{g_3}t^{g_4})^{q_1}{}_{q_2}.
\end{eqnarray}

\Quarklines\, may be created using the \texttt{Quark\_line(std::string)}
constructor. Closed quark-lines are denoted by standard parenthesis,
whereas curly brackets represent open quark-lines. For example, we may write
\begin{eqnarray}
  & &\Quarkline\; \texttt{Ql1("(Nc\^{}2-1)(1,2,3,4)");}\nonumber\\
  & &\Quarkline\; \texttt{Ql2("\texttt{Nc TR \{1,3,4,2\}}");}  
\end{eqnarray}
for the above quark-lines.
Among the \quarkline\, member functions we especially note the
functions for contracting neighboring and next to neighboring gluon
indices, \texttt{contract\_neighboring\_gluons()} and 
\texttt{contract\_next\_neighboring\_gluons()},
as well as the \texttt{normal\_order()} member function which
orders a closed \Quarkline\, such that the smallest gluon index is 
written first. Thus, for example
\begin{eqnarray}
  & & \Quarkline\; \texttt{Ql("(2,3,4,4,5,1)");}  \nonumber\\
  & & \texttt{Ql.contract\_neighboring\_gluons();}\\
  & & \texttt{Ql.normal\_order();}\nonumber
\end{eqnarray}
results in \texttt{\CFtt(1,2,3,5)}.
The other public member functions are listed in 
\tabref{tab:quarkline} in \appref{sec:member_functions}.

\subsubsection{Col\_str}
\label{sec:colstr}

A general color amplitude consists not only of one quark-line,
but of a linear combination of products of \Quarklines. 
One term in this linear combination, a \Polynomial\, times a
product of closed and open quark-lines is contained in a \Colstr, 
\begin{equation}
  \begin{tabular}{ccccc}
    \Colstr & $\sim$ & \Polynomial & $\times$  & \colstr\\
    & & \texttt{Poly}       & $\times$  & \texttt{cs}\\
  \end{tabular}.
\end{equation}
Here the actual information about the color structure is 
carried by a \colstr, (technically a vector of 
\Quarklines), for example we may thus have 
\begin{eqnarray}
  \texttt{(Nc\^{}2-1) \{1,3,4,2\}(5,6)(7,8) } \sim (\Nc^2-1)\; (t^{g_3} t^{g_4})^{q_1}{}_{q_2} \tr(t^{g_5}t^{g_6})\; \tr( t^{g_7}t^{g_8}).
\end{eqnarray}
Like the other color classes, \Colstrs\, may be created using a 
string constructor
\begin{eqnarray}
  & &\Colstr\; \texttt{Cs("(Nc\^{}2-1)[\{1,3,4,2\}(5,6)(7,8)]");} 
\end{eqnarray}
where we note that the \colstr\, is written inside square brackets.
The \Polynomial\, should multiply the whole \colstr, rather than 
individual \quarklines.

The \Colstr\, member functions, listed in \tabref{tab:colstr},
overlap to a high degree with the \Quarkline\, member functions.
In particular there are functions for contracting neighboring 
and next to neighboring gluons, and a 
\texttt{normal\_order} member function.
Apart from normal ordering the individual \Quarklines, the \Colstr\,
\texttt{normal\_order()} member function sorts the \Quarklines\,
as described in \tabref{tab:colstr} in \appref{sec:member_functions},
where the public members of \Colstr\, are listed.

For \Colstr\, (and \Quarkline\, and \Colamp), there is also a \texttt{simplify()}
member function. This function removes quark-lines with 0 quarks,
(these are \Nctt\, for closed \Quarklines\, and \texttt{1} for open 
\Quarklines), normal orders the \colstrs, and simplifies the 
\Polynomials. Potential \Polynomials\, multiplying individual 
\Quarklines\, are also moved to the \Colstr\, member variable 
\texttt{Poly}.

\subsubsection{Col\_amp}
\label{sec:colamp}

The linear combination of \Colstrs\, which in general
is needed to keep track of a color structure is contained
in a \Colamp. In order to contain scalar results, i.e.,
terms not containing any color structure (which arise 
in the process of color contraction) a \Colamp\,
also contains a \Polynomial\, part, \texttt{Scalar},
\begin{equation}
  \begin{tabular}{ccccccc}
  \Colamp & $\sim$ & \texttt{Polynomial} &+& \colamp \nonumber \\
          &        & \texttt{Scalar} &+& \texttt{ca} \nonumber \\
  \mbox{where} & &\\
  \texttt{ca} & $\sim$ & $\Colstr_0$+ $\Colstr_1$+$\Colstr_2$+ ...\nonumber \\
  \end{tabular}.
\end{equation}
For example, we may have the \Colamp
\begin{eqnarray}
  \texttt{(1,2)(3,4) - Nc\^{}{(-1)}(1,2,3,4)} 
  \sim \tr(t^{g_1}t^{g_2})\; \tr( t^{g_3}t^{g_4}) -\frac{1}{\Nc} \tr(t^{g_1}t^{g_2}t^{g_3}t^{g_4}), 
\end{eqnarray}
where \texttt{Scalar} is \texttt{0}, and the \Polynomials\, 
multiplying the \Colstrs\, \texttt{(1,2)(3,4)} and \texttt{(1,2,3,4)} 
are \texttt{1} and \texttt{1/Nc} respectively.

Also the \Colamp\, class has a string constructor which reads in to the
\colamp\, part according to the syntax
\begin{eqnarray}
  & &\Colamp\; \texttt{Ca("[(1,2,3,4)]-1/Nc[(1,2)(3,4)]");}.
\end{eqnarray}
As for \Quarkline\, and \Colstr, \Colamp\, has member functions 
for contracting all neighboring and next to neighboring gluons.
Gluon contraction between gluons which are further away on the
\Quarkline\, may result in a sum of \Quarklines.
Such contractions cannot be seen as actions on a single
\quarkline\, and are therefore not implemented in the
\Quarkline\, and \Colstr\, classes. 
Instead the \Colamp\, class contains functions 
for contracting a gluon or contracting all gluons in the \Colamp.
These functions are intended for \Colamps\, with closed \Quarklines, i.e.,
quarks should be contracted first, using 
\texttt{contract\_quarks(\Colamp1, \Colamp2)}.
When calculating 
a scalar product, all quark indices are thus contracted first,
followed by all the gluon indices. The complete list of 
public member functions is given in \tabref{tab:colamp}
in \appref{sec:member_functions}.

\subsection{Col\_functions -- a library class}
\label{sec:colfunctions}

\Colfunctions\, is a library class containing functions 
which cannot, in a natural way, be attributed to one class, 
or functions which act on many classes and are therefore  
conveniently collected into one class.

In particular, \Colfunctions\, contains classes for 
evaluating scalar products, for numerical evaluation, 
for taking the leading $\Nc$ limit, and for 
describing the effect of gluon emission or
gluon exchange.

\subsubsection{Numerical evaluation}
\label{sec:numerical}
\Colfunctions\, is the class which carries numerical values 
of $\Nc$, $\TR$ and $\CF$, stored in the private members
\Nctt, \TRtt\, and \CFtt\, respectively. 
To numerically evaluate a \Monomial\, to a \texttt{double} or a 
complex number (\texttt{cnum})
the function \texttt{cnum\_num} or \texttt{double\_num} is used 
\begin{eqnarray}
    & &\texttt{\Colfunctions\; Col\_fun;}\nonumber \\
    & &\texttt{\Monomial\; Mon;}\nonumber \\
    & &\texttt{Col\_fun.cnum\_num(Mon);}\\
    & &\texttt{Col\_fun.double\_num(Mon);}\nonumber
\end{eqnarray}
The syntax for numerical evaluation of \Polynomials, \Polyvecs\,
and \Polymatrs\, is identical.

\subsubsection{Leading $\Nc$ evaluation}
\label{sec:leading}

\Colfunctions\, also contains functions for taking the leading $\Nc$
limit of the classes \Polynomial, \Polyvec\, and \Polymatr.
The leading $\Nc$ terms can be evaluated in two different ways,
either by taking the strict $\Nc \to \infty$ limit and dropping all
color suppressed terms (default), or by keeping the color 
suppressed terms in $\CF=\TR(\Nc^2-1)/\Nc$, while
dropping other color suppressed terms. 
For keeping the full $\CF$ in numerical evaluations the 
member variable \texttt{full\_CF} must be set to true 
using \texttt{set\_full\_CF(true)}.

\subsubsection{Scalar products}
\label{sec:scalarproducts}

Scalar products are evaluated using the \Colfunctions\,
member functions \texttt{scalar\_product},  
\begin{eqnarray}
  & &\texttt{\Colstr\; Cs1("[\{1,3,4,2\}(5,6)]");}\nonumber \\
  & &\texttt{\Colstr\; Cs2("[\{1,3,4,5,6,2\}]");}\nonumber \\
  & &\texttt{\Colamp\; Ca1(Cs1);}\nonumber \\
  & &\texttt{\Colamp\; Ca2(Cs2);}\nonumber \\
  & &\texttt{Col\_fun.scalar\_product(Ca1,Ca2);}\\
  & &\texttt{Col\_fun.scalar\_product(Cs1,Cs2);}\nonumber
\end{eqnarray}
in both cases resulting in $\TR\Nc \CF^3$.

\subsubsection{Gluon exchange and gluon emission}

For soft gluon resummation, for NLO calculations, and 
for the cancellation of real emissions and virtual corrections 
in the soft limit, the color structure associated with gluon 
exchange plays an important role.
The \Colfunctions\, class therefore has functions for describing
the effect of gluon exchange on a \Colstr\, or on a \Colamp. 
For example, we may exchange a gluon between parton \texttt{3} and \texttt{6} in 
\texttt{Cs1} 
\begin{eqnarray}
  & &\texttt{Col\_fun.exchange\_gluon(Cs1,3,6);}
\end{eqnarray}
 resulting in the \Colamp
\begin{eqnarray}
\texttt{-TR[\{1,5,6,3,4,2\}]+TR[\{1,3,5,6,4,2\}]+TR[\{1,6,5,3,4,2\}]-TR[\{1,3,6,5,4,2\}]}.\nonumber
\end{eqnarray}
To understand the signs we note that each time a gluon is inserted before
the emitter on a (closed or open) quark-line there is a minus sign, and each time
a gluon is inserted after, there is a plus sign. We also 
note that in this case, the result of gluon exchange on a 
single color structure gave rise to a linear combination 
of four color structures, the maximal possible number of color 
structures from a single \colstr\, \cite{Sjodahl:2009wx}.

In the context of gluon exchange we also remark that \Colfunctions\, can
calculate the ``color correlator'', $\left<\Col|\mathbf{T}_i \cdot \mathbf{T}_j|\Col\right>$ arising 
when coherently emitting a gluon from parton $i$ and $j$ in an amplitude
$\left|\Col\right>$, or when exchanging a gluon between the partons $i$ and $j$ in $\left|\Col\right>$. 
For example we can calculate the color correlator for exchanging a 
gluon between parton \texttt{1} and 
\texttt{4} in \texttt{Ca1},
\begin{eqnarray}
  & &\texttt{Col\_fun.color\_correlator(Ca1,1,4);}
\end{eqnarray}
resulting in a \Polynomial\, with value 
$\TR^3 \CF^{2} + \TR^2 \Nc \CF^3$.

Sometimes, for example in the context of a parton shower, 
one may be interested in the effect of gluon emission itself. 
Starting in a \Colamp\, \texttt{Ca1} and emitting a gluon,
\texttt{7}, from parton \texttt{3} this may be found using
\begin{eqnarray}
  & &\texttt{Col\_fun.emit\_gluon(Ca1,3,7);}
\end{eqnarray}
giving the \Colamp\,
\texttt{[\{1,3,7,4,2\}(5,6)]-[\{1,7,3,4,2\}(5,6)]}.

\subsection{Classes for color bases}
\label{sec:basisclasses}

Although \ColorFull\, may perform calculations with individual 
\Quarklines, \Colstrs\, and \Colamps, the intended usage
is via the color basis classes \Tracebasis\, (in particular), 
\Treelevelgluonbasis\, and \Orthogonalbasis.

As these classes share much of the most important functionality,
they all inherit from one base class, \Colbasis. 
The \Tracebasis\, and \Treelevelgluonbasis\, classes inherit 
from \Colbasis\, via \Tracetypebasis, whereas \Orthogonalbasis\,
inherits directly from \Colbasis.

\subsubsection{Col\_basis}

The \Colbasis\, class has a \colbasis\, member variable 
\texttt{cb} for containing the basis vectors,
\begin{eqnarray}
  \begin{tabular}{cc}
    \texttt{Col\_basis}& $\sim$ \texttt{ col\_basis}\\
    & \texttt{cb}
  \end{tabular}.
\end{eqnarray}
\Colbasis\, also carries
information about the number of quarks and the number of gluons, 
in the public members \texttt{nq} and \texttt{ng}, and (if it has been calculated)
the scalar product matrix in polynomial form, \texttt{P\_spm},
and in double form, \texttt{d\_spm}, as well as the
leading (see \secref{sec:leading}) scalar product matrix in polynomial
and double versions, \texttt{leading\_P\_spm} and 
\texttt{leading\_d\_spm}.

The most important \Colbasis\, member function is probably 
the \texttt{scalar\_product\_matrix()} function which 
calculates the matrix of scalar products between the basis 
vectors. In its default from, this function uses memoization,
as this speeds up the calculations, but this may be 
circumvented in using the 
\texttt{scalar\_product\_matrix\_no\_mem()} version.

In order not to have to calculate scalar product matrices 
over and over again \Colbasis\, also has functions 
for reading in and writing out scalar product matrices,
both in numerical and polynomial form. 

\Colbasis\, also contains functions for reading in and writing out
the basis itself. This is essential for the \Orthogonalbasis\,
class which cannot (presently) construct the orthogonal bases, 
but may read them in.

The decomposition of color amplitudes into bases is done
with the (virtual) \texttt{decompose(\Colamp)} member function,
which does the decomposition by exploring the coefficients 
in front of traces and products of traces for \Tracebasis\,
and \Treelevelgluonbasis\, and by evaluating scalar products 
for the \Orthogonalbasis\, case.

Another important function is the \texttt{color\_gamma} function
which (using \texttt{decompose}) calculates the soft anomalous dimension matrix,
i.e., calculates the matrix describing the effect of gluon exchange
between two partons. The result is returned as a \texttt{Poly\_matr} where 
component $i,j$ gives the amplitude for ending up in vector $i$,
if starting in vector $j$, see also section \secref{sec:standalone}.
A list of public members and functions for \Colbasis\, is found in
\tabref{tab:colbasis}.

\subsubsection{Trace\_type\_basis}

\Tracetypebasis\, is a small helper class for keeping track of 
functions which are similar for \Tracebasis\, and \Treelevelgluonbasis. 
It inherits from \Colbasis\, and is inherited from by
\Tracebasis\, and \Treelevelgluonbasis. 
Most importantly, this is where \texttt{decompose} is implemented 
for these two classes, see \tabref{tab:tracetypebasis}.

\subsubsection{Trace\_basis}

Although the observation that each color amplitude may be decomposed 
into products of open and closed quark-lines is a guiding principle 
for \ColorFull, the \Tracebasis\, class itself is rather small,
containing mainly functions for creating bases. 

A trace basis is created by first contracting all $\qqbar$-pairs 
in all $\Nq!$ ways, and then attaching gluons to these closed
quark-lines, and to additional closed quark-lines, such that at 
least two gluons are attached to each closed quark-line.  

This can be done either directly using a constructor 
\begin{equation}
  \Tracebasis\; \texttt{Tb(2,2);}
\end{equation}
or as
\begin{eqnarray}
  & &\Tracebasis\; \texttt{Tb;}\nonumber\\
  & &\texttt{Tb.create\_basis(2,2)};
\end{eqnarray}
The result, which can be written out to a user given or
default file 
(\texttt{write\_out\_Col\_basis}) or to \texttt{cout}
(using \texttt{<<}) has the basis vectors
\begin{equation}
\begin{tabular}{cc}
\texttt{0} &    \texttt{ [\{1,5,6,2\}\{3,4\}]}\\
\texttt{1} &  \texttt{   [\{1,5,6,4\}\{3,2\}]}\\
\texttt{2} &  \texttt{   [\{1,6,5,2\}\{3,4\}]}\\
\texttt{3} &  \texttt{   [\{1,6,5,4\}\{3,2\}]}\\
\texttt{4} &   \texttt{    [\{3,5,6,2\}\{1,4\}]}\\
\texttt{5} &   \texttt{    [\{3,5,6,4\}\{1,2\}]}\\
\texttt{6} &   \texttt{    [\{3,6,5,2\}\{1,4\}]}\\
\texttt{7} &   \texttt{    [\{3,6,5,4\}\{1,2\}]}\\
\texttt{8} &    \texttt{   [\{1,5,2\}\{3,6,4\}]}\\
\texttt{9} &   \texttt{    [\{1,5,4\}\{3,6,2\}]}\\
\texttt{10}&    \texttt{    [\{1,6,2\}\{3,5,4\}]}\\
\texttt{11}&    \texttt{    [\{1,6,4\}\{3,5,2\}]}\\
\texttt{12}&    \texttt{    [\{1,2\}\{3,4\}(5,6)]}\\
\texttt{13}&    \texttt{    [\{1,4\}\{3,2\}(5,6)]}
\end{tabular}
\label{eq:22}
\end{equation}
At tree-level, in pure QCD, the last two basis vectors vanish.
To create a basis which is only valid up to order \texttt{n\_loop}
in pure QCD, we may use the \texttt{create\_basis(n\_quark,n\_gluon, n\_loop)}
member function. See \tabref{tab:tracebasis} for member functions.

\subsubsection{Tree\_level\_gluon\_basis}

In the case of gluon-only color structures, charge conjugation
implies that each trace must appear with its conjugate, in
linear combinations of the form 
$\tr[t^{g1}t^{g2}...t^{gn}] + (-1)^{\Ng}\tr[t^{gn}...t^{g2}t^{g1}]$. 
In the trace type basis class 
\Treelevelgluonbasis\, this is used to reduce the number of 
basis vectors and speed up calculations. More information 
can be found in \tabref{tab:treelevelgluonbasis}.

\subsubsection{Orthogonal\_basis}

\ColorFull\, can not -- in its present form -- automatically 
create multiplet bases.
However, orthogonal bases may be read in using the 
(\Colbasis) function \texttt{read\_in\_Col\_basis(std::string)}
member function.

For dealing with orthogonal bases, \ColorFull\, offers 
special functions for calculating the matrix of 
scalar products and decomposing vectors. 
The syntax for basis reading is the same as for basis writing.
Bases thus appear much as as in \eqref{eq:22}, 
see \tabref{tab:colbasis} and \tabref{tab:orthogonalbasis}.

\section{Validation}
\label{sec:validation}

For a code with order 10 000 lines, validation is essential.
For this reason \ColorFull\, is continuously validated
using a test suite, which aims at testing all the various 
components. The applied tests starts with checking basic
functions for reading in and writing out files, and 
dealing with polynomials. After this, the creation of bases
is tested, and scalar products are tested by changing 
the order of index contraction, and by switching on and 
off memoization. The scalar product matrices have further 
been tested against \ColorMath\, \cite{Sjodahl:2012nk} 
and, for processes occurring in the context of 
\cite{Campanario:2013fsa}, also against another private 
Mathematica code. The functions describing gluon emission
and gluon exchange are cross-checked against each other.

\section{Conclusion and outlook}
\label{sec:conclusions}

\ColorFull, a C++ stand-alone QCD color algebra package, 
designed for interfacing with event generators, has been 
presented.

\ColorFull\, is based on trace bases, which can automatically 
be created, and color contraction is performed by 
repeated usage of the Fierz identity. Employing these bases, 
one can in principle describe any QCD process.
In reality, the scalar product matrices, which
may be calculated once and for all, become hard to 
manage for more than approximately 8 gluons plus 
$\qqbar$-pairs.
This limitation is inherent for the trace bases, since they
are non-orthogonal, and for this reason \ColorFull\, is 
written to be able to load and use orthogonal 
(multiplet) bases.  

\ColorFull\, does, however, not -- in its present form --
perform index contraction in terms of group invariants
as described in for example \cite{Cvi08, DecompositionPaper}. 
Extending \ColorFull\, to inherently construct orthogonal
multiplet bases and efficiently perform index 
contraction using $3j$ and $6j$ coefficient may speed
up the treatment of QCD color space very significantly.

\section*{Acknowledgments}

Simon Pl\"atzer is thanked for interfacing \ColorFull\, to
\Herwig\, and for writing the build system.
I also want to thank Simon Pl\"atzer and Johan Gr\"onqvist for 
many valuable discussions on the organization of this code.
This work was supported by a Marie Curie Experienced Researcher fellowship of 
the MCnet Research Training network, contract MRTN-CT-2006-035606, 
by the Helmholtz Alliance "Physics at the Terascale" and by the 
Swedish Research Council, contract number 621-2010-3326, 621-2012-27-44
and 621-2013-4287.

\clearpage
\appendix
\section{Class member functions}
\label{sec:member_functions}

\begin{table}[h]
  \caption{\label{tab:monomial} 
    Public data members and functions of the class \Monomial.}
  \begin{tabular}[t]{ | p{7cm} |  p{8.4cm} |}
    \hline\hline

    \bf Data members & \bf Content
    \\\hline\hline

    \texttt{cnum cnum\_part} &
    Complex number multiplying the monomial.
    \\\hline

    \texttt{int int\_part}&
    Integer multiplying the monomial, can be 0.
    \\\hline

    \texttt{int pow\_CF}&
    Power of \texttt{CF=TR*(Nc\^{}2-1)/Nc}.
    \\\hline
    
    \texttt{int pow\_Nc}&
    Power of the number of colors.
    \\\hline
    
    \texttt{int pow\_TR}&
    Power of \texttt{TR} in the \texttt{Monomial}.
    \\\hline\hline
        
    \bf Constructors & \bf Effect
    \\\hline\hline

    \texttt{Monomial()} &	
    Default constructor that sets \texttt{int\_part=cnum\_part=1}, and 
    \texttt{pow\_Nc=pow\_TR=pow\_CF=0.}	
    \\\hline

    \texttt{Monomial(double dnum)} &
    Constructor using a \texttt{double}.
    The \texttt{cnum\_part} member is set to contain the value.
    \\\hline

    \texttt{Monomial(int num)}   & 
    Constructor using an \texttt{int}.
    The \texttt{int\_part} member is set to contain the value.
    \\\hline

    \texttt{Monomial(std::string str)}&
    Constructor taking a string as argument.
    The argument should be of the form in for example
    ``\texttt{-(20*TR\^{}5)/Nc}'' or ``\texttt{-20 TR\^{}(5)/Nc}'' or ``\texttt{20 / TR\^{}(-5)Nc\^{}(1) CF\^{}(3)}''.
    {\it Note}: All spaces and all \texttt{*} are ignored, 
    except in  \texttt{*(-1)} and \texttt{*-1}, 
    which are understood as \texttt{*(-1)}.
    {\it Everything} standing after \texttt{/} is
    divided out, whereas everything standing before \texttt{/} is multiplied with.
    Parentheses are ignored unless they appear in powers, directly after \^{}.
    No spaces are allowed inside the powers.
    If the string contains no information or is empty, the \texttt{Monomial} is set to 1,
    i.e., \texttt{pow\_TR = pow\_Nc = pow\_CF = 0, int\_part = 1, cnum\_part = 1.0}.
    (Expanded Mathematica expressions are of this form.)
    \\\hline\hline

    \bf Member functions & \bf Effect
    \\\hline\hline

    \texttt{void conjugate()}  &
    Take the complex conjugate.
    Note that this changes the \texttt{Monomial} itself.
    \\\hline

    \texttt{void read\_in\_Monomial(std::string filename)}&
    Function for reading in the \texttt{Monomial} from the 
    file \texttt{filename}. The syntax is as for 
    \texttt{Monomial(std::string)}, and comments starting
    with \texttt{\#} are allowed at the top of the file.
    \\\hline
    
    \texttt{void write\_out\_Monomial(std::string filename) const}&    
    Function for writing out the \texttt{Monomial} to a file
    with name \texttt{filename}.
    \\\hline

  \end{tabular}
\end{table}

\begin{table}[h]
  \caption{\label{tab:polynomial} 
    Public data members and  functions of the class \Polynomial.}
  \begin{tabular}[t]{ | p{7cm} |  p{8.4cm} |}
    \hline\hline

    \bf Data member & \bf Content
    \\\hline\hline
    \hspace*{-3 mm}
    \begin{tabular}{l}
    \texttt{polynomial poly}\\ 
    where \texttt{polynomial} is \\
    \texttt{std::vector<Monomial>}
    \end{tabular}& \vspace*{-0.7 cm}
    Contains the \polynomial\,, a sum of \texttt{Monomials}.
    An empty \polynomial\, is defined as 1, to get 0, multiply with 0.
    \\\hline\hline

    \bf Constructors & \bf Effect
    \\\hline\hline

    \texttt{Polynomial()} &
    Default constructor, leaves \polynomial\, empty (=1).
    \\\hline
    
    \texttt{Polynomial(const std::string str)}&
    Constructor allowing setting the \texttt{Polynomial} by using a string,
    should be used as for example: "\texttt{Polynomial Poly("(-20*TR\^{}(5))/Nc + 28*Nc*TR\^{}(5) - 10*Nc\^{}3*TR\^{}(5)")}".
    The \Monomials\, should be separated by \texttt{+} or \texttt{-}, see also the
    \Monomial\, string constructor.
    \\\hline
    
    \texttt{Polynomial(double dnum)}&    
    Constructor allowing setting the \Polynomial\, using a double.
    The \Polynomial\, gets one \Monomial\, where the real part of
    \texttt{cnum\_part} equals \texttt{dnum}.
    \\\hline

    \texttt{Polynomial(int num)}&    
    Constructor allowing setting the \texttt{Polynomial} using an 
    \texttt{int}. The \texttt{Polynomial} gets one 
    \texttt{Monomial} where \texttt{int\_part} has the value \texttt{num}.
    \\\hline\hline

    \bf Member functions & \bf Effect
    \\\hline\hline

    \texttt{void append(const Monomial Mon)}&
    Adding a \texttt{Monomial} term.
    \\\hline

    \texttt{const Monomial\& at(int i) const}&
    Returns the \texttt{Monomial} at place \texttt{i}.
    \\\hline

    \texttt{Monomial\& at(int i)}&    
    Returns the \texttt{Monomial} at place \texttt{i}.
    \\\hline

    \texttt{void clear()}&    
    Erases the information in \texttt{polynomial}.
    \\\hline
    
    \texttt{void conjugate()}&
    Takes the complex conjugate of the \polynomial.
    Note that this changes the \texttt{Polynomial} itself.
    \\\hline

    \texttt{bool empty()}&   
    Is the \texttt{polynomial} empty?
    \\\hline

    \texttt{void erase(int i)}&    
    Erase the \texttt{Monomial} at place \texttt{i}.
    \\\hline
    
    \texttt{void normal\_order()}&    
    Orders terms in \texttt{Polynomial} in a unique form,
    first according to \texttt{pow\_Nc}+\texttt{pow\_CF}, 
    then according to \texttt{pow\_Nc} (for same \texttt{pow\_Nc}+\texttt{pow\_CF})
    then according to \texttt{int\_part}*\texttt{cnum\_part}, then according to 
    \texttt{int\_part}, and finally according to \texttt{pow\_TR}.
    \\\hline

    \texttt{void read\_in\_Polynomial(std::string filename)}&
    Function for reading in the \texttt{Polynomial} from the file \texttt{filename}.
    Comments starting with \texttt{\#} are allowed at the
    top of the file.
    \\\hline

    \texttt{void remove\_CF()}&
    Replaces \texttt{CF} with \texttt{TR*Nc-TR/Nc}.
    \\\hline

    \texttt{void simplify()}&
    Collects terms with the same power of \texttt{TR}, \texttt{Nc} and \texttt{CF}.
    \\\hline

    \texttt{int size()}&    
    Returns the number of terms in the \texttt{Polynomial}.
    \\\hline

    \texttt{void write\_out\_Polynomial( std::string filename) const} &    
    Function for writing out the \texttt{Polynomial} to a file
    with name \texttt{filename}.
    \\\hline
  \end{tabular}
\end{table}

\begin{table}[h]
  \caption{\label{tab:polyvec} 
    Public members and functions in the class \Polyvec.}
  \begin{tabular}[t]{ | p{7cm} |  p{8.4cm} |}
    \hline\hline

    \bf Data member & \bf Content
    \\\hline\hline

    \hspace*{-3 mm}
    \begin{tabular}{l}
    \texttt{poly\_vec pv}\\ 
    where \polyvec\, is:\\
    \texttt{typedef std::vector<Polynomial>}
    \end{tabular}
    &\vspace*{-0.7 cm}
    To actually contain the vector of \Polynomials.
    \\\hline\hline

    \bf Constructors & \bf Effect
    \\\hline\hline

    \texttt{Poly\_vec()}&
    Default constructor, leaves \texttt{pv} empty.
    \\\hline

    \texttt{Poly\_vec(poly\_vec poly\_v)}&
    Makes a \Polyvec\, of a \polyvec.
    \\\hline\hline

    \bf Member functions & \bf Effect
    \\\hline\hline

    \texttt{void append(Polynomial Poly)} &
    Appends a \Polynomial\, to data member \texttt{pv}.
    \\\hline

    \texttt{const Polynomial\& at(int i) const} &
    Returns the \Polynomial\, at place \texttt{i}.
    \\\hline

    \texttt{Polynomial\& at(int i)} &
    Returns the \Polynomial\, at place \texttt{i}.    
    \\\hline

    \texttt{void clear()} &
    Erases the information in the vector.
    \\\hline

    \texttt{void conjugate()}&
    Conjugates the \Polyvec.    
    \\\hline

    \texttt{bool empty() const} &
    Is the vector empty?
    \\\hline

    \texttt{void normal\_order()}&
    Normal orders all \Polynomials\, in the \polyvec\, member \texttt{pv}
    (uses the \texttt{Polynomial::normal\_order} function).
    \\\hline

    \texttt{void read\_in\_Poly\_vec(std::string filename)}&
    Reads in a \Polynomial\, vector of the form \texttt{\{Poly1, Poly2,...}\}
    to the member \texttt{pv} from the file \texttt{filename}.
    Comments starting with \texttt{\#} are allowed
    at the top of the file.
    \\\hline

    \texttt{void remove\_CF()}&
    Remove \texttt{CF} in the \texttt{poly\_vec} member \texttt{pv}, i.e., 
    replace \texttt{CF} by \texttt{TR*Nc-TR/Nc}.
    \\\hline

    \texttt{void simplify()}&
    Simplifies all polynomials in the \polyvec\, member \texttt{pv}
    (uses the \texttt{simplify} member function in \Polynomial).    
    \\\hline
    
    \texttt{uint size() const} &
    Returns the number of \Polynomials\, in the vector,
    i.e., the size of the member \texttt{pv}.
    \\\hline

    \texttt{void write\_out\_Poly\_vec(std::string filename) const}&
    Writes out the vector to the file \texttt{filename}.
    \\\hline
    
  \end{tabular}
\end{table}

\begin{table}[h]
  \caption{\label{tab:polymatr} 
    Public members and functions of the class \Polymatr.}
  \begin{tabular}[t]{ | p{7cm} |  p{8.4cm} |}
    \hline\hline

    \bf Data member & \bf Content
    \\\hline\hline

    \hspace*{-3 mm}
    \begin{tabular}{l}
    \texttt{poly\_matr pm}\\ 
    where \polymatr\, is:\\
    \texttt{typedef std::vector<Poly\_vec>}\\
    \end{tabular}&\vspace*{-0.7 cm}
    To actually contain the matrix of \Polynomials.
    \\\hline\hline

    \bf Constructor & \bf Effect
    \\\hline\hline

    \texttt{Poly\_matr()}&
    Default constructor, leaves \texttt{pm} empty.
    \\\hline\hline

    \bf Member functions & \bf Effect
    \\\hline\hline
    
    \texttt{Poly\_vec\& at(int i)}&
    Returns the \Polyvec\, at place \texttt{i}.
    \\\hline

    \texttt{const Poly\_vec\& at(int i)}&
    Returns the \Polyvec\, at place \texttt{i}.
    \\\hline

    \texttt{Polynomial\& at(int i, int j)}&
    Returns the matrix element at \texttt{i, j}.
    \\\hline
    
    \texttt{const Polynomial\& at(int i, int j)}&
    Returns the matrix element at \texttt{i, j}.
    \\\hline
    
    \texttt{void append(Poly\_vec Pv)}&
    Appends a \texttt{Poly\_vec} to the data member \texttt{pm}.
    \\\hline
    
    \texttt{void clear()}&
    Erases the matrix information.
    \\\hline

    \texttt{void conjugate()}&
    Conjugates the matrix.
    \\\hline
    
    \texttt{bool empty()}&
    Is the matrix, stored in \texttt{pm}, empty?
    \\\hline

    \texttt{void normal\_order()}&
    Normal orders all \polynomials\, in the \polymatr\, member \texttt{pm}
    (uses \texttt{Polynomial::normal\_order}.)
    \\\hline

    \texttt{void read\_in\_Poly\_matr(std::string filename)}&
    \hspace*{-3 mm}
    \begin{tabular}{l}
    Reads in the matrix from the file filename.\\
    The file should be of the format\\
    \texttt{\{\{Poly11,...,Poly1n\},}\\
    \texttt{ ...,}\\
    \texttt{\{Polyn1,...,Polynn\}\}},\\
    and may contain comment lines starting with \texttt{\#} \\
    at the top.
    \end{tabular}
    \\\hline

    \texttt{void remove\_CF()}&
    Removes \texttt{CF} in the \polymatr\, member \texttt{pm}, i.e., replaces \texttt{CF} by
    \texttt{TR*Nc-TR/Nc}.
    \\\hline

    \texttt{void simplify()}&
    Simplifies all \polynomials\, in the \texttt{poly\_matr} member \texttt{pm},
    (uses \texttt{Polynomial.simplify}).
    \\\hline

    \texttt{uint size()}&
    Returns the size of the matrix, the number of \Polyvecs\,
    in the member \texttt{pm}.
    \\\hline
    
    \texttt{void write\_out\_Poly\_matr( std::string filename) const}&
    Writes out the matrix to the file \texttt{filename}.
    \\\hline

  \end{tabular}
\end{table}

\begin{table}[h]
  \caption{\label{tab:quarkline} 
    Public members and functions of the class \Quarkline.}
  \begin{tabular}[t]{ | p{7cm} |  p{8.4cm} |}
    \hline\hline
    
    \bf Data members & \bf Content
    \\\hline\hline

    \hspace*{-3 mm}    
    \begin{tabular}{l}
      \texttt{quark\_line ql}\\
      where \quarkline\, is: \\
      \texttt{typedef std::vector<int>}\\ 
    \end{tabular}&\vspace*{-0.7 cm}
    To actually contain the color information,
    in order $\{q, g_1,g_2,...g_n, \qbar\}$
    or $(g_1, g_2,...g_n)$.
    \\\hline

    \texttt{bool open}&
    Is the string open, with a quark in the beginning and an antiquark in the end, or not?
    \\\hline

    \texttt{Polynomial Poly}&
    \Polynomial\, factor, multiplying the \quarkline.
    \\\hline\hline

    \bf Constructors & \bf Effect
    \\\hline\hline

    \Quarkline( ) &
     Default constructor, leaves \texttt{ql} empty.
    \\\hline

    \Quarkline(\texttt{const std::string}) &
    \hspace*{-3 mm}
    \begin{tabular}{l}
      Constructor used to set the color structure\\ 
      using a string. The string should be of form\\
      \Polynomial*\quarkline, used as \\
      \Quarkline\, \texttt{Ql("5*TR*Nc\^{}2 \{1,6,7,2\}")};, \\
      for an open \Quarkline\, with a quark with\\ 
      number \texttt{1}, two gluons with number \texttt{6} and \texttt{7}, \\
      and a $\qbar$ with number \texttt{2}.\\ 
      For a closed \Quarkline\, the syntax is \\
      \Quarkline\, \texttt{Ql("(1,2,3)");}\\
      The integers should be positive. The \Polynomial\, \\
      should be in a format which it is readable by\\ 
      the \Polynomial\texttt{(std::string)} constructor. 
    \end{tabular}
    \\ \hline\hline

    \bf Member functions & \bf Effect
    \\\hline\hline

    \texttt{\Quarkline\, after(int j) const} & Returns a \Quarkline\, where the \texttt{ql} member is changed to contain only partons after place \texttt{j}.
    \\\hline
    
    \texttt{void append(int p)}
    & Appends parton \texttt{p} to the \Quarkline.
    \\\hline
    
    \texttt{void append(std::vector<int> in\_ql)} &
    Appends a whole \quarkline\, to the \Quarkline. 
    
    \\\hline
    \texttt{int at(int j) const} & 
    Returns the parton at place \texttt{j}. For closed \quarklines\, 
    \texttt{j} may be between -size and 2*size.
    \\\hline

    \texttt{\Quarkline\, before(int j) const} & Returns a \Quarkline\, where the \texttt{ql} member is changed to contain only partons before place \texttt{j}.
    \\\hline
    
    \texttt{void clear()}&
    Erase the information in \quarkline\, \texttt{ql}.
    \\\hline
    
    \texttt{void conjugate()}&
    Conjugates the \Quarkline\, by reversing the \quarkline\, \texttt{ql} 
    and conjugating the \Polynomial\, \texttt{Poly}.
    \\\hline
    
    \texttt{void contract\_neighboring\_gluons(int j)} &
    Contracts neighboring gluons in the \Quarkline\, starting at \texttt{j},
    only intended for closed \Quarklines.
    \\\hline
    
    \texttt{void contract\_neighboring\_gluons() }&
    Function for contracting neighboring gluons in a \Quarkline\, starting at place 0,
    and looking everywhere, only intended for closed \Quarklines.
    \\\hline
  \end{tabular}
\end{table}
\begin{table}
  \begin{tabular}[t]{ | p{7cm} |  p{8.4cm} |}
    \hline
 
    \texttt{void contract\_next\_neighboring \_gluons(int j) }&
    Contracts next to neighboring gluons in the \Quarkline,
    starting at place \texttt{j} (i.e. checking first gluon \texttt{j} and \texttt{j+2}).
    Also looks for new neighbors, only intended for closed \Quarklines.
    \\\hline

    \texttt{void contract\_next\_neighboring \_gluons() }&
    Contracts neighboring and next to neighboring gluons in the \Quarkline,
    starting with contracting neighbors, only intended for closed \Quarklines.
    \\\hline
    
    \texttt{bool empty() const} &
    Is the \quarkline\, empty?
    \\\hline
    
    \texttt{void erase(int i)} &
    To erase the parton at place \texttt{i}.
    \\\hline
    
    \texttt{void insert(int j, int p)} &
    Inserting parton \texttt{p} at place \texttt{j}.
    \\\hline
    
    \texttt{void normal\_order()}&
    Orders a closed \quarkline, such that the smallest gluon index stands first
    (i.e., use that the trace is cyclic).
    \\\hline
 
    \texttt{void prepend(int p)} & 
    Prepends parton \texttt{p} to the \Quarkline.
    \\\hline

    \texttt{void prepend(std::vector<int> in\_ql)} &
    Prepends a whole \quarkline\, to the \Quarkline.
    \\\hline

    \texttt{void read\_in\_Quark\_line( std::string filename )} &
    Function for reading in the \Quarkline\, from the file \texttt{filename}.
    The syntax for the \Quarkline\, is the same as for the \Quarkline\,
    string constructor. 
    \\\hline

    \texttt{uint size() const} &
    The size of the \quarkline.
    \\ \hline
    \texttt{int smallest(const \Quarkline \& Ql1, const \Quarkline \& Ql2) const} &
    Function for finding the "smallest" \Quarkline\, of \texttt{Ql1} and \texttt{Ql2},
    used for deciding which \Quarkline\, should stand first when normal ordering \Colstrs.
    This function does {\it not} first normal order the \Quarklines.
    If only one is open, that \Quarkline\, should stand first.
    If both are open or both are closed, the longest \Quarkline\, should stand first.
    If the size is the same, the \Quarkline\, with smallest starting number should stand first.
    If the first number is the same, check the second number, then the third etc.
    \texttt{1} is returned if \texttt{Ql1} should stand first, and \texttt{2} 
    if \texttt{Ql2} should stand first.
    If \texttt{Ql1==Ql2}, \texttt{0} is returned.
    \\ \hline

    \texttt{std::pair<\Quarkline, \Quarkline> split\_Quark\_line(int j1, int j2) const} &
    Function for splitting a closed \Quarkline\, into two \Quarklines.
    The gluons at \texttt{j1} and \texttt{j2} are removed in the split.
    May create 1-rings and 0-rings.
    \\ \hline

    \texttt{void write\_out\_Quark\_line( std::string filename) const} &
    Function for writing out the \Quarkline\, to a file with name \texttt{filename}.
    \\ \hline
  \end{tabular}
\end{table}

\clearpage
\begin{table}[h]
  \caption{\label{tab:colstr} 
    Public data members an functions of the class \Colstr.}
  \begin{tabular}[t]{ | p{7cm} |  p{8.4cm} |}
    \hline\hline

    \bf Data members & \bf Content
    \\\hline\hline
    \hspace*{-3 mm}
    \begin{tabular}{l}
      \texttt{col\_str cs}\\ 
      where \colstr\, is:\\
      \texttt{typedef std::vector<Quark\_line>}\\ 
    \end{tabular}&\vspace*{-0.7 cm}
    For containing the information about the color structure,
    a product of \Quarklines,
    contained in a vector of \quarklines.
    \\ \hline
    
    \texttt{Polynomial Poly}&
    \Polynomial\, factor multiplying the whole product of \Quarklines.
    \\\hline\hline
        
    \bf Constructors & \bf Effect
    \\\hline\hline

    \texttt{\Colstr()}&
    Default constructor, leaves \texttt{cs} empty.
    \\ \hline
    
    \texttt{\Colstr(\Quarkline\, Ql)}&
    Make a \Colstr\, of a \Quarkline.
    \\ \hline

    \texttt{\Colstr(const std::string str)}&
    \hspace*{-3 mm}
    \begin{tabular}{l}
    Constructor for setting the color structure using\\
    a string. Should be used as:\\
    \texttt{Cs("Nc*TR\^{}(3) [\{1,2,3,4\}(5,6)(7,8)]")},\\
    i.e., the argument should be\\ 
    \Polynomial\, * \colstr.\\
    (The \Polynomial\, should multiply the whole \\ 
    \colstr, i.e., stand outside the \texttt{[]}-brackets.
    \end{tabular}
    \\ \hline\hline

    \bf Member functions & \bf Effect
    \\\hline\hline

    \texttt{const \Quarkline \& at(int i)}&
    Returns the \Quarkline\, at place \texttt{i}.
    \\ \hline

    \texttt{\Quarkline \& at(int i)}&
    Returns the \Quarkline\, at place \texttt{i}.
    \\ \hline  
    
    \texttt{int at(int i, int j) const}&
    Returns the parton at place \texttt{j} in in \Quarkline\, \texttt{i}.
    \\ \hline
    
    \texttt{void append(Quark\_line Ql)}&
    Appends a \Quarkline\, to the data member \texttt{cs}.
    \\ \hline

    \texttt{void append(col\_str cs\_in)}&
    Append the content of a \colstr\, to the \texttt{cs} of the \Colstr.
    \\ \hline
    
    \texttt{void clear()}&
    Erase information in \colstr.
    \\ \hline

    \texttt{void conjugate()}&
    Function for conjugating the \Colstr\, by conjugating each \Quarkline\, in \texttt{cs},
    as well as the \Polynomial\, \texttt{Poly}.
    \\ \hline
    
    \texttt{void contract\_2\_rings()}&
    Function for contracting gluon indices in closed \Quarklines\, with only 2 gluons.
    This removes the 2-ring, replaces one of the gluon indices, and
    multiplies with a factor $\tr[t^a t^a]=\TRtt$ (no sum),
    only intended for fully contractable \Colstrs.
    \\ \hline
    \texttt{void contract\_next\_neighboring \_gluons( )}&
    Contracts neighboring and next to neighboring gluons in each
    \Quarkline\, in the \Colstr, starting with contracting neighbors.
    This function should only be used on \Colstrs\, with only closed \Quarklines.
    \\ \hline

    \texttt{void contract\_quarks(}
    \texttt{const \Colstr\, Cs1, const \Colstr\, Cs2)}
    &
    Function for contracting quarks between two color structures \texttt{Cs1} and \texttt{Cs2}.
    The result is stored in the \Colstr\, itself.
    \\ \hline
    \texttt{bool empty() const}&
    Is the \colstr\, empty?
    \\ \hline
  \end{tabular}
\end{table}
\begin{table}
  \begin{tabular}[t]{ | p{7cm} |  p{8.4cm} |}
    \hline

    \texttt{void erase(int i)}&
    Erases the \Quarkline\, at place \texttt{i}.
    \\ \hline
    \texttt{void erase(int i, int j)}&
    Erases the parton at position \texttt{j} in \Quarkline\, \texttt{i}.
    \\ \hline
    \texttt{void erase(std::pair<int, int> place)}&
    Erases the parton located at \texttt{place}.
    \\ \hline

    \texttt{std::string find\_kind(int p) const}&
    Finds out if a parton is a quark, anti-quark or gluon,
    returns "q", "qbar" or "g" respectively.
    This function does {\it not} loop over all partons, but assumes
    that the parton is a gluon if it is in a closed \Quarkline,
    or if the \Quarkline\, is open but the parton cannot be found in the ends.
    \\ \hline
    
    \vspace*{-3 mm} \hspace*{-3 mm}
    \begin{tabular}{l}
      \texttt{std::pair<int, int>}\\
      \texttt{find\_parton(int part\_num) const}
    \end{tabular}&
    Locates the parton with number \texttt{part\_num} in a \colstr.
    \\ \hline
    
    \texttt{bool gluons\_only() const}&
    Checks if the amplitude only has gluons, i.e., if all \Quarklines\, are closed.
    \\ \hline

    \texttt{void insert(int i, int j, int part\_num)}&
    To insert the parton \texttt{part\_num} in \quarkline\, \texttt{i}
    at place \texttt{j}.
    \\ \hline 
    \texttt{bool left\_neighbor(int p1, int p2) const}&
    Function for telling if parton \texttt{p2} stands to the left of parton \texttt{p1}.
    \\ \hline
    
    \texttt{int longest\_quark\_line() const}&
    Returns the length of the longest \Quarkline\, in the \colstr.
    \\ \hline

    \texttt{int n\_gluon() const}&
    Counts the number of gluons in a \Colstr.
    Counts all gluon indices, both free and contractable.
    \\ \hline

    \texttt{int n\_quark() const}&
    Counts the number of quarks (=number of anti-quarks) in a \Colstr.
    Counts all quark indices, both free and contracted.
    \\ \hline

    \texttt{bool neighbor(int p1, int p2) const}&
    Function for telling if the partons \texttt{p1} and \texttt{p2} are neighbors.
    \\ \hline

    \texttt{void normal\_order()}&
    Normal orders the \Colstr\, by first
    normal ordering individual \Quarklines\,
    and then normal ordering different \Quarklines\, in \texttt{cs}.
    For the ordering, see the member function \texttt{smallest} in this
    class and in the \Quarkline\, class.
    \\ \hline

    \texttt{void read\_in\_Col\_str(std::string filename)}&
    Function for reading in the \Colstr\, from the file \texttt{filename}.
    \\ \hline

    \texttt{void remove\_0\_rings()}&
    Removes \Quarklines\, without partons, equal to \Nctt\, (closed) or 1 (open).
    \\ \hline

    \texttt{void remove\_1\_rings()}&
    Removes \Quarklines\, with only one gluon as $\tr(t^a)=0$.
    \\ \hline

    \texttt{void replace(int old\_ind, int new\_ind)}&
    Replaces the parton index \texttt{old\_ind} with \texttt{new\_ind}.
    \\ \hline

    \texttt{bool right\_neighbor(int p1, int p2) const}&
    Function for telling if parton \texttt{p2} stands to the right of parton \texttt{p1}.
    \\ \hline

  \end{tabular}
\end{table}
\begin{table}
  \begin{tabular}[t]{ | p{7cm} |  p{8.4cm} |}

    \hline
    \texttt{void simplify()}&
    Removes 0- and 1-rings,
    moves factors multiplying the individual \Quarklines\, to
    multiply the \colstr\, instead (i.e., being stored in \texttt{Poly}),
    simplifies the \Polynomial\, and normal orders the \quarklines.
    \\ \hline
    
    \texttt{uint size()}&
    The size of the \colstr, number of \quarklines.
    \\ \hline
    
      \texttt{int smallest(const \Colstr \& Cs1,} 
      \texttt{const \Colstr \& Cs2) const}
      &
    Finds out the "smallest" \Colstr\, of two \Colstrs, i.e.,
    which \Colstr\, should stand first in a normal ordered \Colamp\, or basis.
    Returns \texttt{1}, if \texttt{Cs1} should stand before \texttt{Cs2}
    and \texttt{2} if \texttt{Cs2} should stand before \texttt{Cs1}.
    Both \Colstrs\, have to be normal ordered for the result to be unique.
    The \Colstrs\, are ordered by
    (1) number of \Quarklines\,
    (2) if the \Quarkline\, at place 0,1,2... is open or not
    (3) the size of the \Quarkline\, at place 1,2,3...
    (4) the parton numbers in the \Quarklines\, at place 1,2,3...,
    i.e., first the first parton in the first \Quarkline\, is checked
    and last the last parton in the last \Quarkline.
    The function returns \texttt{0} if \texttt{Cs1==Cs2}.
    \\ \hline

    \texttt{void write\_out\_Col\_str(}
    \texttt{std::string filename) const}
    &
    Function for writing out the \Colstr\, to a file
    with name \texttt{filename}.\\
\hline
\end{tabular}
\end{table}

\clearpage
\begin{table}[h]
  \caption{\label{tab:colamp} 
    Public members and functions of the class \Colamp.}
  \begin{tabular}[t]{ | p{7cm} |  p{8.4cm} |}
    \hline\hline

    \bf Data members & \bf Content
    \\\hline\hline

    \hspace*{-3 mm}
    \begin{tabular}{l}
      \texttt{col\_amp ca}\\ 
      where \colamp\, is:\\
      \texttt{typedef std::vector<Col\_str>}\\ 
    \end{tabular}&\vspace*{-0.7 cm}
    To actually contain the information about the \Colstrs, 
    \texttt{ca=Cs1+Cs2+Cs3+...} .
    \\ \hline

    \texttt{Polynomial Scalar}&
    \texttt{Scalar} is a \Polynomial\, for collecting color factors appearing 
    when the color structure has been fully contracted.
    The full color amplitude is \texttt{Scalar+Cs1+Cs2+Cs3...}. 
    \texttt{Scalar} should thus be non-zero 
    only if all indices can be contracted.
    \\\hline\hline

    \bf Constructors & \bf Effect
    \\\hline\hline
    
    \texttt{\Colamp()}&
    Default constructor, sets \texttt{Scalar = 0}, and 
    leaves \texttt{ca} empty.
    \\ \hline
    
    \texttt{\Colamp(Col\_str Cs)}&
    Constructor converting a \Colstr\, to a \Colamp.
    \\ \hline
    
    \texttt{\Colamp(const std::string str)}&
    \hspace{-3 mm}
    \begin{tabular}{l}
      Constructor taking a string as argument.\\
      The string should be of the form\\
      \Polynomial1*\colstr1+\Polynomial2*\colstr2,\\
      for example:\\
      \texttt{Ca("Nc*[(1,3,4,2)]+1/Nc [(1,4)(3,2)]")}.\\
      (The \Polynomials\, should multiply the whole\\ 
      \colstrs, and thus stand outside the \texttt{[]}-brackets.) 
    \end{tabular}
    \\ \hline\hline

    \bf Member functions & \bf Effect
    \\\hline\hline
    
    \texttt{void append(\colamp\, ca\_in)}&
    Appends the \Colstrs\, in \texttt{ca\_in} to the \colamp\, member \texttt{ca}.
    \\ \hline
    
    \texttt{const Col\_str \& at(int i)}&
    Returns the \Colstr\, at place \texttt{i}.
    \\ \hline
  
    \texttt{Col\_str \& at(int i)}&
    Returns the \Colstr\, at place \texttt{i}.
    \\ \hline
    
    \texttt{void clear()}&
    Erases the information about the color amplitude, stored in \texttt{ca}.
    \\ \hline
   
    \texttt{void collect\_col\_strs()}&
    Compares \colstrs\, in the \Colamp\, to collect similar \colstrs\,
    and only store once in \texttt{ca}.
    \\ \hline
    
    \texttt{void conjugate()}&
    Function for taking the conjugate of the \Colamp\,
    by conjugating each \Colstr\, in \texttt{ca} and the
    \Polynomial\, member \texttt{Scalar}.
    \\ \hline
    
    \texttt{void contract\_2\_rings()}&
    Contract closed \Quarklines\, with only 2 gluons in
    each \Quarkline\, in each \Colstr\, in the \Colamp.
    This removes the 2-ring, replaces one of the gluon indices and
    multiplies with a factor $\tr[t^a t^a]=\TRtt$ (no sum),
    only intended for fully contractable \Colamps.
    \\ \hline
    
    \texttt{void contract\_a\_gluon()}&
    Contracts one gluon, the first gluon in the first \Quarkline\, 
    (in each \Colstr), only intended for closed \Quarklines.
    \\ \hline

  \end{tabular}
\end{table}
\begin{table}
  \begin{tabular}[t]{ | p{7cm} |  p{8.4cm} |}
    \hline
     
    \texttt{void contract\_all\_gluons()}&
    Function for contracting all gluon indices in a \Colamp,
    only intended for closed \Quarklines.
    \\ \hline
    
    \texttt{void contract\_next\_neighboring \_gluons()}&
    Contracts up to next to neighboring gluons in each \Quarkline\,
    in each \Colstr\, in each \Colamp, only intended for closed \Quarklines.
    \\ \hline

      \texttt{void contract\_quarks(const \Colamp\,} 
      \texttt{Ca1, const \Colamp\, Ca2)}
      &
    Function for contracting the (anti-)quarks in \texttt{Ca1} with those
    in \texttt{Ca2}. The results is saved in the \Colamp\, itself.
    \\ \hline
    
    \texttt{void contract\_Quark\_line\_gluons()}&
    Function for contracting gluon indices within the \Quarklines.
    Checks only for {\it one} pair in each \Quarkline.
    \\ \hline

    \texttt{bool empty() const}&
    Is the \colamp\, \texttt{ca} empty?
    \\ \hline
    
    \texttt{void erase(int i)}&
    Erases the \Colstr\, at place \texttt{i}.
    \\ \hline
    
    \texttt{bool gluons\_only() const}&
    Checks if the \Colamp\, only contains gluons, i.e., if all \Quarklines\, are closed.
    \\ \hline
    
    \texttt{int longest\_quark\_line() const}&
    Returns the length of the longest \Quarkline\, in any \Colstr.
    \\ \hline
    
    \texttt{int n\_gluon() const }&
    Returns the number of gluons in the \Colamp\, as the number of gluons in the first \Colstr.
    Note that the other \Colstrs\, could have a different number of (contracted) gluons.
    (Intended for tree-level \Colamps\, with only one \Colstr.)
    \\ \hline

    \texttt{int n\_gluon\_check() const} &
    Returns the number of gluons in the \Colamp\, after checking that each \Colstr\,
    has the same number of gluons.
    \\ \hline 

    \texttt{int n\_quark() const}&
    Returns the number of quarks  in the \Colamp\, as the number of quarks in the first \Colstr.
    Note that the other \Colstrs\, could have a different number of (contracted) quarks.
    (Intended for tree-level \Colamps\, with only one \Colstr.)
    \\ \hline

    \texttt{int n\_quark\_check() const}&
    Returns the number of quarks in the \Colamp\, after checking that each \Colstr\,
    has the same number of quarks.
    \\ \hline

    \texttt{void normal\_order()}&
    Normal orders the individual \colstrs\, and then
    orders the \Colstrs\, using the order defined in
    the \Colstr\, member function \texttt{smallest}.
    \\ \hline  
    
    \texttt{void normal\_order\_col\_strs()}&
    Normal orders all \colstrs\, in \texttt{ca}.
    \\ \hline
    
      \texttt{void read\_in\_Col\_amp(}
      \texttt{std::string filename)}
    &
    Reads in the \Colamp\, to the member \texttt{ca} from the file \texttt{filename}.
    (This is intended for reading in an actual color amplitude,
    nothing is read in to the \texttt{Polynomial} member \texttt{Scalar}.)
    Comments starting with \texttt{\#} are allowed at the top of the file.
    \\ \hline
    
    \texttt{void remove\_0\_rings()}&
    Removes \quarklines\, with no gluons, these are \Nctt\, if closed, and defined to be 1 if open.
    \\ \hline
  \end{tabular}
\end{table}

\begin{table}
  \begin{tabular}[t]{ | p{7cm} |  p{8.4cm} |}
    \hline
    \texttt{void remove\_1\_rings()}&
    Removes \Colstrs\, with \quarklines\,  with only 1 gluon, these are 0 as $\tr[t^a]=0$.
    \\ \hline
    
    \texttt{void remove\_empty\_Col\_strs()}&
    Removes empty \Colstrs. An empty \Colstr\, means that all indices have been contracted,
    so the \Colstr\, is equal to its \Polynomial, which is moved to the \texttt{Scalar} part
    of the \Colamp.
    \\ \hline

    \texttt{void simplify()}&
    Function for simplifying an amplitude.
    Removes 0 and 1-rings,
    compares \colstrs,
    removes \Colstrs\, multiplying 0, and
    simplifies \Polynomials\, of the individual \Colstrs.
    \\ \hline

    \texttt{uint size() const}&
    The size of the \colamp\, \texttt{ca}.
    \\ \hline
      \texttt{void write\_out\_Col\_amp(}
      \texttt{std::string filename) const}
      &
    Function for writing out the \Colamp\, to a file
    with name \texttt{filename}.
    \\ \hline
  \end{tabular}
\end{table}

\clearpage
\begin{table}[h]
\caption{\label{tab:colfunctions} 
Some (private) data members and the  
public functions of the library class \Colfunctions.}
\begin{tabular}[t]{ | p{7cm} |  p{8.4cm} |}
  \hline\hline
 
  \bf Data members (private) & \bf Content
  \\\hline\hline
  
  \texttt{double CF}&
  The value of $\CF=\TR (\Nc^2-1)/\Nc$,
  changed by the \texttt{set\_CF} function.
  Note that \texttt{CF} can be changed independently of \texttt{Nc}.
  \\ \hline

  \texttt{bool full\_CF}&
  While finding the leading terms in a \Polynomial\, one may want to keep the full value of 
  \texttt{CF}, \texttt{TR*Nc-TR/Nc}, 
  or only keep the leading \texttt{Nc} term \texttt{TR*Nc} (default).
  The switch \texttt{full\_CF} is used by the \Polynomial\, version of \texttt{leading}
  (and hence also by the \Polyvec\, and \Polymatr\, versions etc.).
  The \texttt{leading} functions replaces \texttt{CF} by \texttt{TR*Nc} if 
  \texttt{full\_CF} is \texttt{false} (default)
  while evaluating the \texttt{leading} terms.
  If \texttt{full\_CF} is \texttt{true}, \texttt{CF} is replaced by \texttt{TR*Nc-TR/Nc}.
  Clearly this affects the result of subsequent numerical evaluation.
  In the \Colbasis\, class (and derived) the matrix version of \texttt{leading}
  is used to evaluate scalar product matrices.
  \\ \hline

  \texttt{double Nc}&
  The number of colors, used in numerical results,
  changed by the \texttt{set\_Nc} function.
  \\ \hline
  
  \texttt{double TR} &
  The trace convention $\tr( t^a t^a )$=TR (no sum),  
  the normalization of the SU($\Nc$) generators, to be used in numerical evaluation.
  This value can be changed by the \texttt{set\_TR} function.
  \\ \hline\hline

  \bf Constructor & \bf Effect
  \\\hline\hline

  \texttt{\Colfunctions()}&
  Default constructor, sets \texttt{\Nctt=3, \TRtt=0.5, \CFtt=4.0/3.0} and
  \texttt{full\_CF=false}.
  \\ \hline\hline

  \bf Member functions & \bf Effect
  \\\hline\hline

  \texttt{cnum cnum\_num(const Monomial\& Mon) const}&
  Numerically evaluates a \Monomial\, using the \Nctt, \TRtt\, and \CFtt\, variables.
  \\ \hline

  \texttt{cnum cnum\_num(const Polynomial\& Poly) const}&
  Numerically evaluates a \Polynomial, using the \Nctt\, \TRtt\, and \CFtt\, variables.
  \\ \hline
  
  \texttt{cvec cnum\_num(const Poly\_vec\& Pv) const}&
  Numerically evaluates a \Polyvec\, (vector of \Polynomials),
  using \texttt{cnum\_num(\Polynomial)}.
  \\ \hline

  \texttt{cmatr cnum\_num(const Poly\_matr\& Pm) const}&
  Numerically evaluates a \Polymatr\, (vector of \Polyvec),
  using \texttt{cnum\_num(\Polyvec)} for each \Polyvec.
  \\ \hline

\end{tabular}
\end{table}
\begin{table}
\begin{tabular}[t]{ | p{7cm} |  p{8.4cm} |}
  \hline

  \texttt{Polynomial color\_correlator(const Col\_amp Ca, int i, int j) const}&
  Calculates $\left< \Col | \mathbf{T}_{i} \mathbf{T}_{j} | \Col \right>$, the "color correlator"
  relevant for coherent gluon emission from
  parton \texttt{i} and parton \texttt{j}, or gluon exchange between \texttt{i} and \texttt{j}.
  The \texttt{Ca} thus corresponds to ${\left| \Col \right>}$, 
  and \texttt{i} and \texttt{j} are the partons involved in the emission (exchange).
  \\ \hline 

  \texttt{double double\_num(const Monomial \& Mon) const}&
  Numerically evaluates a \Monomial\, to a \texttt{double},
  using the data members \Nctt, \CFtt\, and \TRtt.
  \\ \hline
  
  \texttt{double double\_num(const Polynomial \& Poly) const}&
  Numerically evaluates a \Polynomial\, to a \texttt{double}, 
  using the data members \Nctt, \CFtt\, and \TRtt.
  \\ \hline

  \texttt{dvec double\_num(const Poly\_vec \& Pv) const}&
  Numerically evaluates a \Polyvec, vector of \Polynomial.
  \\ \hline
  
  \texttt{dmatr double\_num(const Poly\_matr \& Pm) const}&
  Numerically evaluates a \texttt{Poly\_matr} (vector of \texttt{Poly\_vec)}.
  \\ \hline

  \texttt{Col\_amp emit\_gluon(const Col\_str \& in\_Col\_str, int emitter, int g\_new) const}&
  Function for emitting a gluon from a \Colstr.
  When the gluon is inserted before the emitter in a \Quarkline,
  the amplitude comes with a minus sign.
  \\ \hline
  
  \texttt{Col\_amp emit\_gluon(const Col\_amp \& Ca\_in, int emitter, int g\_new) const}&
  Function for emitting a gluon from a \Colamp.
  When the gluon is inserted before the emitter in a \Quarkline,
  the amplitude comes with a minus sign.
  \\ \hline

  \texttt{Col\_amp exchange\_gluon(const \Colstr \& Cs, int p1, int p2) const}&
  Function for exchanging a gluon between the partons \texttt{p1} and \texttt{p2} in the \Colstr\, \texttt{Cs}.
  When the gluon is inserted before the emitter in a \Quarkline,
  the amplitude comes with a minus sign.
  \\ \hline
  
  \texttt{Col\_amp exchange\_gluon(const Col\_amp \& Ca, int p1, int p2) const}&
  Function for exchanging a gluon between two partons  \texttt{p1} and \texttt{p2} in the \Colamp\, \texttt{Ca}.
  When the gluon is inserted before the emitter in a \Quarkline,
  the amplitude comes with a minus sign.
  \\ \hline

  \texttt{double get\_CF() const}&
  Returns the value of \CFtt.
  \\ \hline

  \texttt{double get\_Nc() const}&
  Returns the number of colors.
  \\ \hline
  
  \texttt{double get\_TR() const}&
  Returns the normalization of the generators,
  $\tr(t^a t^b)=\TRtt\delta^{ab}$.
  \\ \hline
  
  \texttt{bool get\_full\_CF() const}& 
  Returns \texttt{true} if \texttt{full\_CF} is true and \texttt{false} otherwise.
  \\ \hline

  \texttt{int factorial(int i) const}&
  The factorial of an \texttt{int}, 0! is defined as 1.
  \\ \hline

  \end{tabular}
\end{table}
\begin{table}
\begin{tabular}[t]{ | p{7cm} |  p{8.4cm} |}
  \hline

  \texttt{Polynomial leading(const Polynomial \& Poly) const}&
  Takes the leading \Nctt\, terms of a \Polynomial, i.e., keeps the \Monomials\, with the highest
  power of \Nctt\, plus \CFtt. If \texttt{full\_CF} is \texttt{false} (default), 
  \CFtt\, is replaced by \TRtt*\Nctt.
  If \texttt{full\_CF} is \texttt{true} \CFtt\, is replaced by \TRtt*\Nctt-\TRtt/\Nctt.
  \\ \hline

  \texttt{Poly\_vec leading(const Poly\_vec\& Pv) const}&
  Takes the leading part of a \Polyvec.
  Keeps only \Monomials\, with maximal power of \CFtt\, plus \Nctt,
  uses \texttt{leading(const Polynomial \& Poly).}
  If \texttt{full\_CF} is \texttt{false} (default), \CFtt\, is replaced by \TRtt*\Nctt.
  If \texttt{full\_CF} is \texttt{true}, \CFtt\, is replaced by \TRtt*\Nctt-\TRtt/\Nctt.
  Note that taking the leading terms of a \Polyvec\, is not necessarily
  the same as taking the leading terms of each \Polynomial.
  \\ \hline

  \texttt{Poly\_matr leading(const Poly\_matr\& Pm) const}&
  Takes the leading part of a matrix of \Polynomials,
  keeping only those with maximal power of \CFtt\, plus \Nctt.
  If \texttt{full\_CF} is false (default), \CFtt\, is replaced by $\TR\Nc$.
  If \texttt{full\_CF} is true \CFtt\, is replaced by \TRtt*\Nctt-\TRtt/\Nctt.
  Note that taking the leading terms of a \Polymatr\, is not necessarily
  the same as taking the leading terms in each \Polyvec.
  \\ \hline

  \texttt{int leading\_Nc\_pow(const Polynomial \& Poly) const}&
  Function for finding the leading power of \Nctt\, in a \Polyvec,
  i.e., the power of \Nctt\, plus the power of \CFtt.
  \\ \hline

  \texttt{int leading\_Nc\_pow(const Poly\_vec \& Pv) const}&
  Function for finding the leading power of \Nctt\, in a \Polyvec.
  \\ \hline

  \texttt{Polynomial Polynomial\_cnum\_num(const Polynomial \& Poly) const}&
  Numerically evaluates a \Polynomial\, using the value of the data member \Nctt,
  and stores in the format of a \Polynomial\, with only one term with only a numerical part.
  \\ \hline
  
  \texttt{Poly\_vec Poly\_vec\_cnum\_num(const Poly\_vec \& Pv) const}&
  Numerically evaluates a \Polyvec\, (vector of \Polynomial)
  and stores in the form of a \Polyvec, uses \texttt{polynomial\_cnum\_num(Pv.at(p))}
  for each \Polynomial.
  \\ \hline
  
  \texttt{Poly\_matr Poly\_matr\_cnum\_num(const Poly\_matr \& Pm) const}&
  Numerically evaluates a \Polymatr\, (vector of \Polyvec)
  and stores in the form of a \Polymatr.
  \\ \hline   

   \end{tabular}
\end{table}
\begin{table}
\begin{tabular}[t]{ | p{7cm} |  p{8.4cm} |}
  \hline

  \texttt{dvec read\_in\_dvec(std::string filename) const}&
  Reads in a numerical vector and saves it as a \texttt{double} matrix, 
  \texttt{dmatr}. The file should be of the format
  \texttt{\{d1,...,dn\}}, and may contain comment lines starting 
  with \texttt{\#} at the top.
  \\ \hline
  
  \texttt{dmatr read\_in\_dmatr(std::string filename) const}&
  \hspace*{-3 mm}
  \begin{tabular}{l}
    Reads in a numerical matrix and save it as a \\
    \texttt{double} matrix, \texttt{dmatr}. The file should be of \\
    the format\\
    \texttt{\{\{d11,...,d1n\},}\\
    \texttt{ ...,}\\
    \texttt{\{dn1,...,dnn\}\}}\\
    and may contain comment lines starting with \texttt{\#}\\
  at the top.
  \end{tabular}
  \\ \hline
  
  \texttt{Polynomial scalar\_product(const Col\_amp \& Ca1, const Col\_amp \& Ca2) const}&
  Function for calculating the scalar product between two \Colamps.
  \\ \hline
  
  \texttt{Polynomial scalar\_product(const Col\_str \& Cs1, const Col\_str \& Cs2) const}&
  Function for calculating the scalar product between two \Colstrs.
  \\ \hline

  \texttt{void set\_Nc(double n)}& 
  Set the number of colors.
  The value of \CFtt\, is adjusted accordingly.
  \\ \hline

  \texttt{void set\_TR(double tr)}&
  Sets the normalization of the generators.
  The value of \CFtt\, is adjusted accordingly.
  \\ \hline
  
  \texttt{void set\_CF(double cf)}&
  Sets the value of \CFtt.
  The value of \Nctt\, is {\it not} adjusted accordingly.
  \\ \hline

  \texttt{void set\_full\_CF(bool is\_full)}&
  Switch on/off \texttt{full\_CF}.
  \\ \hline

  \texttt{void write\_out\_dmatr(const dmatr \& matr, std::string filename) const}&
  Writes out the double version of a (scalar product) matrix
  to the file \texttt{filename}.
  \\ \hline

  \texttt{void write\_out\_dvec(const dvec \& dv, std::string filename) const}&
  Function for writing out a numerical vector,
  to the file \texttt{filename}.
  \\ \hline

  \end{tabular}
\end{table}

\clearpage
\begin{table}[h]
\caption{\label{tab:colbasis} 
Public members and functions of the class \Colbasis.}
\begin{tabular}[t]{ | p{7cm} |  p{8.4cm} |}
  \hline\hline

    \bf Data members & \bf Content
    \\\hline\hline

    \hspace*{-0.3 cm}
    \begin{tabular}{l}
      \texttt{col\_basis cb}\\
      where \texttt{col\_basis} is:\\
      \texttt{typedef std::vector<Col\_amp>}
    \end{tabular}& 
    \vspace*{-0.7 cm}
    Contains the information about the basis vectors 
    \texttt{cb= vector1, vector2...}.
    \\ \hline
 
    \texttt{Col\_functions Col\_fun}&   
    Contains the set of \Colfunctions\, used.
    \\ \hline
    
    \texttt{dmatr d\_spm}&
    To contain the double version of the scalar product matrix.
    \\ \hline
    
    \texttt{dmatr leading\_d\_spm}&
    To contain the double version of the leading part of the scalar product matrix.
    \\ \hline

    \texttt{Poly\_matr leading\_P\_spm}&   
    To contain the \Polynomial\, version of the leading part of the scalar product matrix.
    \\ \hline
      
    \texttt{int ng}&
    The number of gluons, initially set to \texttt{0} and (possibly)
    changed with \texttt{create\_basis}, or while reading in the basis.
    \\ \hline

    \texttt{int nq}&
    The number of $\qqbar$-pairs, initially set to \texttt{0} and (possibly)
    changed with \texttt{create\_basis}, or while reading in the basis.
    \\ \hline

    \texttt{Poly\_matr P\_spm}&
    To contain the \Polynomial\, version of the scalar product matrix.
    \\ \hline\hline

    \bf Constructor and destructor& \bf Effect
    \\\hline\hline

    \texttt{Col\_basis() }&
    Default constructor, sets \texttt{nq=ng=0}, and the private
    data members \texttt{trace\_basis} = 
    \texttt{tree\_level\_gluon\_basis} = \texttt{orthogonal\_basis} = \texttt{false}.
    \\\hline
  
    \texttt{virtual $\sim$Col\_basis()}&
    Destructor.
    \\ \hline\hline

    \bf Member functions & \bf Effect
    \\\hline\hline
    
    \texttt{\Colamp \& at(const int i) }&
    Returns the \Colamp\, (i.e. basis vector) at place \texttt{i}.
    \\ \hline
    
    \texttt{const \Colamp \& at(const int i) const} &
    Returns the \Colamp\, (i.e. basis vector) at place \texttt{i}.
    \\ \hline
    
    \texttt{void append(\Colamp\, Ca) }&
    Appends a \Colamp\, to the basis stored in \texttt{cb}.
    \\ \hline

    \texttt{std::string basis\_file\_name() const}&
    Returns a standard filename, used for writing out
    the basis to a file.
    \\ \hline
    
    \texttt{void clear()}&
    Erase the basis, stored in \texttt{cb}.
    \\ \hline
    
\end{tabular}
\end{table}

\begin{table}
  \begin{tabular}[t]{ | p{7cm} |  p{8.4cm} |}
    \hline
    
    \texttt{Poly\_matr color\_gamma(int p1, int p2)}&
    Function for calculating the color structure part of the soft anomalous
    dimension matrix. First calculates the effect of gluon exchange on a
    basis vector and then decomposes the result into the basis.
    For this to be possible, the basis must clearly contain all resulting
    basis vectors, meaning for example that it can not be
    used for \Treelevelgluonbasis.
    The function is only available for the \Tracebasis\,
    and the \Orthogonalbasis\, classes.
    \\ \hline
    
    \hspace*{-3 mm}
    \begin{tabular}{l}
      \texttt{virtual Poly\_vec decompose(}\\ 
      \texttt{const Col\_amp \& Ca)} 
    \end{tabular}& 
    \vspace*{-5 mm}
    Each type of color basis has to implement a function for decomposing
    an amplitude in the color basis.
    \\ \hline
    
    \texttt{bool empty() const}&
    Is the \colbasis\, empty?
    \\ \hline

    \hspace*{-3 mm}
    \begin{tabular}{l}
      \texttt{Col\_amp exchange\_gluon(}\\
    \texttt{uint vec, int p1, int p2)} 
    \end{tabular}& 
    \vspace*{-5 mm}
    Function for finding the resulting \Colamp\, after exchanging
    a gluon between parton \texttt{p1} and parton \texttt{p2} in the
    basis vector vec.
    \\ \hline
    
    \texttt{bool  is\_Orthogonal\_basis() const }&
    Is it an \Orthogonalbasis?
    \\ \hline
    
    \texttt{bool  is\_Trace\_basis()}&
    Is it a \Tracebasis?
    \\ \hline
    
    \texttt{bool  is\_Tree\_level\_gluon\_basis() }&
    Is it a \Treelevelgluonbasis?
    \\ \hline 
    
    \texttt{void leading\_scalar\_product\_matrix()}&
    Finds the leading \Nctt\, scalar product matrices,
    \texttt{leading\_P\_spm} and \texttt{leading\_d\_spm}.
    If the polynomial scalar product matrix, \texttt{P\_spm} has
    been calculated, \texttt{P\_spm} is used, otherwise \texttt{P\_spm} is first calculated
    and the leading \Nctt\, limit is then taken of \texttt{P\_spm}.
    \\ \hline
    
    \texttt{int n\_gluon\_check() const }&
    Returns the number of gluons in the \Colbasis\, after
    checking that each Col\_str in each Col\_amp
    has the same number of gluons.
    \\ \hline
    
    \texttt{int n\_quark\_check() const }&
    Returns the number of quarks in the \Colbasis\, after
    checking that each \Colstr\, in each \Colamp\,
    has the same number of quarks.
    \\ \hline

    \texttt{virtual void read\_in\_Col\_basis() }&
    Function for reading in the basis from the default filename
    (see \texttt{basis\_file\_name}).
    \\ \hline

  \end{tabular}
\end{table}

\begin{table}
  \begin{tabular}[t]{ | p{7cm} |  p{8.4cm} |}
    \hline

    \begin{tabular}{l}
      \texttt{virtual void read\_in\_Col\_basis(}\\
      \texttt{std::string filename)}
    \end{tabular}& 
    \hspace*{-0.3 cm}
    \begin{tabular}{l}
      Function for reading in the basis from the file \\
      \texttt{filename}. The basis should be in human readable \\
      format, of form:\\
      
      \texttt{0  [\{1,3,4,2\}]}\\
      \texttt{1  [\{1,4,3,2\}]}\\
      \texttt{2  [\{1,2\}(3,4)]}\\
      
      i.e. first the basis vector number 0,1,2..., etc,\\ 
      then the \Colamp\, corresponding to the basis \\
      vector in question.\\
      The \Colamps\, may consist of several \Colstrs,\\ 
      for example\\
      \texttt{0  [(1,2,3,4)]+[(1,4,3,2)]}\\
      \texttt{1  [(1,2,4,3)]+[(1,3,4,2)]}\\
      \texttt{2  [(1,3,4,2)]+[(1,2,4,3)]}\\
      and each \Colstr\, may also contain a \\
      \Polynomial. (The \Polynomial\, should multiply\\ 
      the whole \colstr, rather than a \quarkline, \\
      i.e., the \Polynomial\, should stand outside the \texttt{[]}\\
      -brackets.)
    \end{tabular}
    \\ \hline
    
    \texttt{void read\_in\_d\_spm( )}
    &
    \hspace*{-3 mm}
    \begin{tabular}{l}
      Reads in a numerical matrix from a file with\\ 
      default filename (see \texttt{spm\_file\_name}) and saves it\\
      as a double matrix, \texttt{dmatr},
      in the member \\ 
      variable \texttt{d\_spm}. The file should be of the format\\
      \texttt{\{\{d11,...,d1n\},}\\
      \texttt{...,}\\
      \texttt{\{dn1,....dnn\}\},}\\
      and may contain comment lines starting with
      \\ \texttt{\#} at the top.
    \end{tabular}
    \\ \hline

    \texttt{void read\_in\_d\_spm(std::string filename)}&
    Reads in a numerical matrix from a file \texttt{filename}
    and saves it in the \texttt{double} matrix, \texttt{dmatr},
    in the member variable \texttt{d\_spm}.
    For format, see
    \texttt{read\_in\_d\_spm( )}.
    \\ \hline
    
    \texttt{void read\_in\_leading\_d\_spm( }
    \texttt{std::string filename)}
    & 
    Reads in a numerical matrix from the file \texttt{filename} 
    and saves it as a double matrix, \texttt{dmatr},
    in the member variable \texttt{leading\_d\_spm}.
    The format should be as for
    \texttt{read\_in\_d\_spm( )}, with \texttt{0} for non-diagonal elements.
    \\ \hline
    
    \texttt{void read\_in\_leading\_d\_spm( )}&
    Reads in a numerical matrix from a file with default filename (see \texttt{spm\_file\_name})
    and saves it as a \texttt{double} matrix, \texttt{dmatr}, in the member variable \texttt{leading\_d\_spm}.
    The format should be as for
    \texttt{read\_in\_d\_spm( )}, with 0 for non-diagonal elements.
    \\ \hline

  \end{tabular}
\end{table}
\begin{table}
  \begin{tabular}[t]{ | p{7cm} |  p{8.4cm} |}
    \hline

    \texttt{void read\_in\_leading\_P\_spm(  ) }&
    Reads in a \Polynomial\, matrix from default filename (see \texttt{spm\_file\_name})
    and saves it as a \Polymatr\, in the member variable \texttt{leading\_P\_spm}.
    The format should be as for \texttt{void read\_in\_P\_spm()},
    with \texttt{0} for non-diagonal elements.
    \\ \hline
    
      \texttt{void read\_in\_leading\_P\_spm(} 
      \texttt{std::string filename) }
    &
    Reads in a \Polynomial\, matrix from a file with default filename
    (see \texttt{spm\_file\_name})  and saves it as a \Polymatr\,
    in the member variable \texttt{leading\_P\_spm}.
    The format should be as for \texttt{void read\_in\_P\_spm()},
    with \texttt{0} for non-diagonal elements.
    \\ \hline

    \texttt{void read\_in\_P\_spm( )}&
    \hspace*{-3 mm}
    \begin{tabular}{l}
      Reads in a \Polynomial\, matrix from a file with \\
      default filename (see \texttt{spm\_file\_name}) and saves \\
      it as a \Polymatr\, in the member variable \texttt{P\_spm}. \\
      The file should be in the format\\
      \texttt{\{\{Poly11,...,Poly1n\}},\\
      \texttt{...,}\\
      \texttt{\{Polyn1,...,Polynn\}\}},\\
      and may contain comment lines starting with \texttt{\#} \\
      at the top.
    \end{tabular}
    \\ \hline
    
    \texttt{void read\_in\_P\_spm(std::string filename)}&
    Reads in a \Polynomial\, matrix from the file \texttt{filename} 
    and saves it as a \Polymatr\, in the member variable \texttt{P\_spm}.
    The format should be as for \texttt{void read\_in\_P\_spm()}.
    \\ \hline

      \texttt{void rename\_indices(}
      \texttt{Col\_str \& Cs1, Col\_str \& Cs2) const}
      &
    A function to rename the indices in two \Colstrs, 
    such that in the first  they are called 1,2,3..., 
    and in the second the relative order is kept.
    \\ \hline

    \texttt{virtual Polynomial scalar\_product(}
    \texttt{const Col\_amp \& Ca1, const Col\_amp \& Ca2)} 
    &
    Function for calculating scalar products algebraically
    using the basis and the scalar product matrix (\Polymatr\,  \texttt{P\_spm}) 
    in the basis.
    (Does add implicit conjugated part for \Treelevelgluonbasis,
    as these terms are contained in the matrix of scalar products.)
    \\ \hline
  
    \texttt{void scalar\_product\_matrix() }&
    Function for calculating the scalar products matrix.
    This function loops over all basis vectors and stores the
    value of the scalar product between basis vector
    \texttt{i} and basis vector \texttt{j} in the \texttt{i,j} 
    -entry in \texttt{P\_spm} and \texttt{d\_spm}.
    The calculation is done using memoization.
    The symmetry of the scalar product matrix is not used
    for the calculation, instead it is checked that the
    resulting matrix is indeed symmetric.
    \\ \hline
  
  \end{tabular}
\end{table}
\begin{table}
  \begin{tabular}[t]{ | p{7cm} |  p{8.4cm} |}
    \hline

    \texttt{void scalar\_product\_matrix\_num()}&
    As \texttt{scalar\_product\_matrix}, but does the
    calculation numerically. It hence only calculates \texttt{d\_spm}.
    \\ \hline
    
    \texttt{virtual void scalar\_product\_matrix\_no\_mem()}&
    As \texttt{scalar\_product\_matrix}, but without memoization.
    \\ \hline
    
    \hspace*{-3 mm}
    \begin{tabular}{l}
      \texttt{virtual cnum scalar\_product\_num(} \\
      \texttt{const Col\_amp \& Ca1,} \\
      \texttt{const Col\_amp \& Ca2) }  
    \end{tabular}&
    \vspace*{-7 mm}
    Function for calculating scalar products numerically, knowing the basis
    and the scalar product matrix in numerical form.
    (Does add implicit conjugated part for \Treelevelgluonbasis,
    as these terms are contained in the matrix of scalar products.)
    \\ \hline

    \hspace*{-3 mm}    
    \begin{tabular}{l}
      \texttt{virtual cnum scalar\_product\_num(} \\ 
      \texttt{const cvec \& v1, const cvec \& v2 )}  
    \end{tabular}
    &
    \vspace*{-5 mm}    
    Calculates the scalar product between numerical (complex) amplitudes \texttt{v1}, \texttt{v2}
    using the numerical scalar product matrix, \texttt{d\_spm}.
    The vectors thus needs to be expressed in the basis contained in \texttt{cb}.
    (Does add implicit conjugated part for \Treelevelgluonbasis,
    as these terms are contained in the matrix of scalar products.)
    \\ \hline
    
    \texttt{void scalar\_product\_matrix \_num\_no\_mem()}&
    As \texttt{scalar\_product\_matrix\_num}, but without
    memoization.
    \\ \hline
    
    \texttt{cnum scalar\_product\_num\_diagonal( }
    \texttt{const cvec \& v1, const cvec \& v2) }    
    &
    Calculates the scalar product between numerical (complex) amplitudes \texttt{v1}, \texttt{v2}
     using the numerical scalar product matrix, \texttt{d\_spm}.
     Assumes that there are only diagonal contributions.
     This is useful for calculations in leading \texttt{Nc} limit.
     (Does add implicit conjugated part for \Treelevelgluonbasis,
     as these terms are contained in the matrix of scalar products.)
     \\ \hline
  
     \texttt{void simplify() }&
     Simplifies all the basis vectors by using \Colamp.\texttt{simplify()} on the
     individual \Colamps\, in the basis.
     \\ \hline

     \texttt{const uint size() const}&
    Returns the number of basis vectors.
    \\ \hline
    
     \hspace*{-3 mm}
     \begin{tabular}{l}
       \texttt{std::string spm\_file\_name(}\\
       \texttt{const bool leading,} \\
       \texttt{const bool poly) const}
     \end{tabular}
       &
     \vspace*{-8 mm}
     Returns a standard filename, used for writing out
     scalar product matrices.
     If \texttt{leading} is true, "\_l" is appended to the filename.
     If "\texttt{poly}" is true "\texttt{P\_}" is added to the filename, and if it is
     false "\texttt{d\_}", as in \texttt{double}, is added to the filename.
     \\ \hline
     
     \texttt{virtual void write\_out\_Col\_basis() const }&
     Function for writing out the basis to a file
     with default name (see \texttt{basis\_file\_name}).
     \\ \hline
     
       \texttt{virtual void write\_out\_Col\_basis(}
       \texttt{std::string filename) const} 
       & 
     Function for writing out the basis to a file with name \texttt{filename}.
     \\ \hline

  \end{tabular}
\end{table}
\begin{table}
  \begin{tabular}[h]{ | p{7cm} |  p{8.4cm} |}
    \hline

    \hspace*{-3 mm}    
    \begin{tabular}{l}
      \texttt{void write\_out\_d\_spm(}\\
      \texttt{std::string filename) const}
    \end{tabular} & 
    Writes out the \texttt{d\_spm} to file \texttt{filename}.
    \\ \hline

    \texttt{void write\_out\_d\_spm( ) const}&
    Writes out \texttt{d\_spm} to the standard filename, see \texttt{spm\_file\_name}.
    \\ \hline
    
    \hspace*{-3 mm}
    \begin{tabular}{l}
      \texttt{void write\_out\_leading\_d\_spm(}\\
      \texttt{std::string filename) const}
    \end{tabular} & 
    Writes out \texttt{leading\_d\_spm} to the file \texttt{filename}.
    \\ \hline
 
    \texttt{void write\_out\_leading\_d\_spm( ) const}&
    Writes out \texttt{leading\_d\_spm} to the standard filename, see \texttt{spm\_file\_name}.
    \\ \hline
    
    \hspace*{-3 mm}
    \begin{tabular}{l}
      \texttt{void write\_out\_leading\_P\_spm(}\\
      \texttt{std::string filename) const}
    \end{tabular} & 
    Writes out \texttt{leading\_P\_spm} to the file \texttt{filename}.
    \\ \hline
    
    \texttt{void write\_out\_leading\_P\_spm( ) const}& 
    Writes out \texttt{leading\_P\_spm} to the standard filename, see \texttt{spm\_file\_name}.	
    \\ \hline     

    \hspace*{-3 mm}
    \begin{tabular}{l}
      \texttt{void write\_out\_P\_spm(}\\
      \texttt{std::string filename) const}
    \end{tabular} & 
    Writes out \texttt{P\_spm} to the file \texttt{filename}.
    \\ \hline
    
    \texttt{void write\_out\_P\_spm( ) const}&
    Writes out \texttt{P\_spm} to the standard filename, see \texttt{spm\_file\_name}.
    \\ \hline 
  \end{tabular}
\end{table}

\clearpage
\begin{table}[h]
  \caption{\label{tab:tracetypebasis}
    Data member and public functions of the class \Tracetypebasis.
    As \Tracetypebasis\, inherits from \Colbasis\, all the
    public members of \Colbasis\, are also available, 
    see \tabref{tab:colbasis}.
  }
  \begin{tabular}[t]{ | p{7cm} |  p{8.4cm} |}
    \hline\hline
    \bf Data member (protected) & \bf Content
    \\\hline\hline
    
    \texttt{int max\_ql} &
    The maximal number of quark-lines allowed in the basis.
    This is used for constructing bases that only are valid
    up to a certain order in QCD, such that unused information
    need not be carried around.
    \\\hline\hline

    \bf Member functions & \bf Effect
    \\\hline\hline

    \texttt{Trace\_type\_basis():Col\_basis()}&
    Default constructor, uses the \Colbasis\, constructor and sets
    the variable \texttt{max\_ql} to \texttt{0}.
    \\ \hline 
  
    \texttt{Poly\_vec decompose(const Col\_amp\& Ca)}&
    A function for decomposing the color amplitude \texttt{ca} in the basis,
    returning the result as a \Polynomial.
    \\ \hline 
    
    \texttt{std::pair<int, int>  new\_vector\_numbers(const Col\_str \& Cs, int emitter)}&
    Function for finding the new vector numbers in the new basis
    (this basis)
    after radiating a new gluon from the parton \texttt{emitter} in the
    old color structure is \texttt{Cs}. After emission a linear combination
    of new basis vectors is obtained.
    For emission from a quark or an antiquark there is only one resulting color
    structure, and -1 is returned in the place of the absent color structure.
    The second vector, where the new gluon is inserted before the \texttt{emitter},
    comes with a minus sign in the new total amplitude.
    \\ \hline 
    
    \texttt{std::pair<int, int>  new\_vector\_numbers(int old\_num, int emitter, int n\_loop) const}&
    This function is intended for tree-level processes with at most 2 $\qqbar$-pairs.
    It finds the new vector numbers in the basis for $\Nq$+$\Ng$+1 partons
    after radiating a new gluon from the parton \texttt{emitter}.
    This function does not actually use the \texttt{cb}, but only calculates     
    the basis vector number, which makes it much quicker than the general version.
    The old vector has number \texttt{old\_num}, and there were, before emission,
    $\Nq$ quarks (+ $\Nq$ anti-quarks) and $\Ng-1$ gluons.
    For emission from a quark or an antiquark there is only one resulting color structure,
    and \texttt{-1} is returned in the place of the absent color structure.
    The second vector, where the new gluon is inserted before the emitter
    comes with a minus sign in the new total amplitude.
    \\ \hline

\end{tabular}
\end{table}

\clearpage
\begin{table}[h]
  \caption{\label{tab:tracebasis} 
    Public member functions implemented in the class \Tracebasis.
    As \Tracebasis\, inherits from \Tracetypebasis\, all the
    public members of \Tracetypebasis\, and \Colbasis\, are also available, 
    see \tabref{tab:tracetypebasis} and \tabref{tab:colbasis}.
  }
  
  \begin{tabular}[t]{ | p{7cm} |  p{8.4cm} |}
    \hline\hline

    \bf Constructors & \bf Effect
    \\\hline\hline

    \texttt{Trace\_basis()}&
    Default constructor, calls the \Tracetypebasis\, constructor and
    sets \texttt{nq} = \texttt{ng} = 0. 
    \\ \hline 
    
    \texttt{Trace\_basis(int n\_quark, int n\_gluon)}&
    Constructor for creating a trace basis for \texttt{n\_quark} $\qqbar$-pairs
    and \texttt{n\_gluon} gluons.
    \\ \hline 
    
    \texttt{Trace\_basis(int n\_quark, int n\_gluon, int n\_loop)}&   
    Constructor for creating a trace basis for \texttt{n\_quark} $\qqbar$-pairs
    and \texttt{n\_gluon} gluons, keeping only those color structures that
    can appear to order \texttt{n\_loop} in pure QCD.
    (Note: For electroweak interactions more color structures may
    be needed.)    
    \\ \hline \hline 

    \bf Member functions & \bf Effect
    \\\hline\hline

    \texttt{void create\_basis(int n\_q, int n\_g)} &
    Creates a trace basis with basis vectors saved in the \texttt{cb} member.
    Keeps all possible basis vectors, i.e., the basis is valid to any order
    in perturbation theory.
    \\ \hline 

    \texttt{void create\_basis(int n\_q, int n\_g, int n\_loop)} &
    Creates a trace basis with basis vectors saved in the cb member.
    Keeps only basis vectors consisting of at most \texttt{n\_q} plus
    \texttt{n\_loop} \quarklines.
    \\ \hline

  \end{tabular}
\end{table}

\begin{table}[h]
  \caption{\label{tab:treelevelgluonbasis} 
    Public member functions of the class \Treelevelgluonbasis.
    As \Treelevelgluonbasis\, inherits from \Tracetypebasis\,
    which inherits from \Colbasis, all the
    public members of \Colbasis\, and \Tracetypebasis\, are also available, 
    see \tabref{tab:colbasis} and \tabref{tab:tracetypebasis}.
  }
  \begin{tabular}[t]{ | p{7cm} |  p{8.4cm} |}
    \hline\hline

    \bf Constructors & \bf Effect
    \\\hline\hline

    \texttt{Tree\_level\_gluon\_basis()}&
    Default constructor.
    \\ \hline \hline
    
    \texttt{Tree\_level\_gluon\_basis(int n\_g)} &
    Constructor for creating a tree-level gluon basis\, with \texttt{ng} gluons.   
    \\ \hline\hline

    \bf Member functions & \bf Effect
    \\\hline\hline

    \texttt{void create\_basis(int n\_g)} &
    Creates a basis with basis vectors saved in the \texttt{cb} member.
    Each basis vector is a sum of two traces,
    of form $\tr[t^a t^b....t^z] + (-1)^{\Ng} \tr[t^z .... t^b t^a]$.
    The charge conjugated trace
    is implicit, and only one trace is actually
    carried around which speeds up calculations.
    \\\hline

    \texttt{void read\_in\_Col\_basis( )}&
    Function reading in the basis from default name
    (see \texttt{basis\_file\_name} in the \texttt{Col\_basis} class).
    The full basis, including the charge conjugated
    part should be contained in the file. (This is to simplify
    comparison with other programs, such as \ColorMath.)
    \\\hline

    \texttt{void read\_in\_Col\_basis(std::string filename)}&
    Function for reading in the basis from a file.
    The file should contain the whole basis,
    including the charge conjugated part.
    \\\hline

    \texttt{void write\_out\_Col\_basis( ) const}&
    Function for writing out the basis to default name,
    (see \texttt{basis\_file\_name} in the \texttt{Col\_basis} class).
    The full basis, including the charge conjugated
    part is written out. (This is to simplify
    comparison with other programs, such as \ColorMath.)
    \\\hline

    \texttt{void \texttt{write\_out\_Col\_basis(std::string filename)} const}&
    Function for writing out the basis to \texttt{filename},
    (see \texttt{basis\_file\_name} in the \texttt{Col\_basis} class).
    The full basis, including the charge conjugated
    part is written out. (This is to simplify
    comparison with other programs, such as \ColorMath.)
    \\\hline
 \end{tabular}
\end{table}

\begin{table}[h]
  \caption{\label{tab:orthogonalbasis} 
    Public member functions of the class \Orthogonalbasis.
    \Orthogonalbasis\, inherits directly from \Colbasis,
    meaning that public members of \Colbasis\, are also available, 
    see \tabref{tab:colbasis}.
  }
  \begin{tabular}[t]{ | p{7cm} |  p{8.4cm} |}
    \hline\hline

    \bf Data members & \bf Content
    \\\hline\hline
    \texttt{dvec diagonal\_d\_spm}&
    To contain information about scalar products as a \texttt{dvec},
    i.e., entry \texttt{i} is the square of vector \texttt{i}.
    \\ \hline
    
    \texttt{Poly\_vec diagonal\_P\_spm}&
    To contain information about scalar products as a \texttt{Poly\_vec},
    i.e., entry \texttt{i} is the square of vector \texttt{i}.
    \\ \hline\hline
 
    \bf Constructor & \bf Effect
    \\ \hline\hline

    \texttt{Orthogonal\_basis():Col\_basis()}&
    Default constructor, puts private variable \texttt{orthogonal\_basis =true}
    and calls the constructor of \Colbasis.
    \\ \hline\hline

    \bf Member functions & \bf Effect
    \\ \hline\hline
    
    \texttt{Poly\_vec decompose(const Col\_amp \& Ca)}&
    The decomposition of a \Colamp\, in an orthogonal basis is done
    by calculating scalar products and dividing out the norm.
    The norm is evaluated numerically.
    \\ \hline

    \texttt{void diagonal\_scalar\_product\_matrix( bool save\_P\_diagonal\_spm, bool save\_d\_diagonal\_spm, bool use\_mem)}&
    Calculates the diagonal entries in the scalar product matrix,
    and (depending on arguments), saves them to the member variables
    \texttt{diagonal\_P\_spm} and \texttt{diagonal\_d\_spm}.
    This function is used by the \texttt{Orthogonal\_basis} version of
    \texttt{scalar\_product\_matrix}.
    \\ \hline
    
    \texttt{std::string diagonal\_spm\_file\_name( const bool leading, const bool poly) const}&
    Creates a default filename for writing out diagonal scalar products.
    The boolean variable \texttt{leading} should be \texttt{true} if the name is for a leading
    \texttt{Nc} variable. The filename is then modified accordingly.
   \\ \hline
   
   \texttt{Polynomial scalar\_product(const Col\_amp \& Ca1, const Col\_amp \& Ca2)}&
   Function for calculating scalar products
   given the information about the basis and the scalar product matrix in the basis.
   The \Colamps\, are first decomposed using decompose,
   and then squared using the scalar product matrix \texttt{P\_spm}.
   An orthogonal scalar product matrix is assumed.
   \\ \hline
   
   \texttt{void scalar\_product\_matrix()}&
   Calculates the scalar product matrix assuming the basis to be orthogonal.
   Calculates both the \texttt{double} \texttt{(d\_spm)} and the \Polynomial\, 
   (\texttt{P\_spm}) matrices and saves to default filenames.
   \\ \hline
   
   \texttt{cnum scalar\_product\_num(const Col\_amp \& Ca1, const Col\_amp \& Ca2)}&
   Function for calculating scalar products
   given the information about the basis
   and the scalar product matrix in numerical form.
   The \Colamps\, are first decomposed using \texttt{decompose},
   and then squared using \texttt{diagonal\_d\_spm}.
   For \texttt{Orthogonal\_basis} an orthogonal scalar product matrix is assumed.
   \\ \hline
   
  \end{tabular}
\end{table}
\begin{table}
  \begin{tabular}[t]{ | p{7cm} |  p{8.4cm} |}
    \hline
    \texttt{cnum scalar\_product\_num(const cvec \& v1, const cvec \& v2)}&
    Calculates the scalar product between decomposed amplitudes \texttt{v1}, \texttt{v2}
    using the \texttt{diagonal\_d\_spm} diagonal numerical scalar product matrix.
    The vectors needs to be expressed in the basis contained in \texttt{cb},
    i.e., the decomposition has to be known.
    \\ \hline
    
    \texttt{void write\_out\_diagonal\_d\_spm( std::string filename) const}&
    Writes out \texttt{diagonal\_d\_spm} to the file \texttt{filename}.
    \\ \hline
    
    \texttt{void write\_out\_diagonal\_d\_spm( ) const}&
    Writes out \texttt{diagonal\_d\_spm} to the standard filename, 
    see \texttt{diagonal\_spm\_file\_name}.
    \\ \hline
    
    \texttt{void write\_out\_diagonal\_P\_spm( std::string filename) const}&
    Writes out \texttt{diagonal\_P\_spm} to the file \texttt{filename}.
    \\ \hline
    
    \texttt{void write\_out\_diagonal\_P\_spm( ) const}&
    Writes out \texttt{diagonal\_P\_spm} to the standard filename, 
    see \texttt{diagonal\_spm\_file\_name}.
    \\ \hline
    
    \texttt{void write\_out\_diagonal\_spm(const dvec \& dv, const bool leading) const}&
    Function for writing out a \texttt{dvec} (the diagonal scalar products, \texttt{diagonal\_d\_spm})
    to a file with standard filename given by \texttt{diagonal\_spm\_file\_name}.
    The boolean variable \texttt{leading} should be true if \texttt{dv} only has leading
    \texttt{Nc} contributions. The filename is then modified accordingly.
    \\ \hline
    
    \texttt{void write\_out\_diagonal\_spm(const Poly\_vec \& pv, const bool leading) const}&
    Function for writing out a \Polyvec\, (the diagonal scalar products, \texttt{diagonal\_d\_spm})
    to a file with standard filename given by \texttt{diagonal\_spm\_file\_name}.
    The boolean variable \texttt{leading} should be true if \texttt{Poly\_vec only} has leading
    \texttt{Nc} contributions. The filename is then modified accordingly.
    \\ \hline 
\end{tabular}
\end{table}

\clearpage
\bibliographystyle{JHEP}  
\bibliography{Refs}

\clearpage

\end{document}